\documentclass[runningheads]{llncs}

\usepackage{graphicx}
\usepackage{mathpartir}
\usepackage{amsmath}
\usepackage{amssymb}
\usepackage{stmaryrd}
\usepackage{xcolor}
\usepackage{bbold}
\usepackage[capitalise,noabbrev]{cleveref}
\usepackage{bm}
\usepackage{newunicodechar}
\usepackage[all]{xy}
\usepackage{tikz-cd}

\newunicodechar{◂}{} 

\newcommand\superequiv{\mathrel{\rlap{\raisebox{\fontdimen22\textfont2}{$=$}}\raisebox{-0.5\fontdimen22\textfont2}{$ = $}}}

\newunicodechar{×}{\ensuremath{\times}}
\newunicodechar{→}{\ensuremath{\mathnormal\to}}
\newunicodechar{←}{\ensuremath{\mathnormal\leftarrow}}
\newunicodechar{⟦}{\ensuremath{\mathnormal\llbracket}}
\newunicodechar{⟧}{\ensuremath{\mathnormal\rrbracket}}
\newunicodechar{ℕ}{\ensuremath{\mathbb{N}}}
\newunicodechar{⊞}{\ensuremath{\mathnormal\boxplus}}
\newunicodechar{⟨}{\ensuremath{\mathnormal\langle}}
\newunicodechar{⟩}{\ensuremath{\mathnormal\rangle}}
\newunicodechar{∪}{\ensuremath{\mathnormal\cup}}
\newunicodechar{⦃}{\ensuremath{\mathnormal\{\mskip-4.5mu\mid}}
\newunicodechar{⦄}{\ensuremath{\mathnormal\mid\mskip-4.5mu\}}}
\newunicodechar{⊎}{\ensuremath{\mathnormal\uplus}}
\newunicodechar{✴}{\ensuremath{\mathnormal\ast}}
\newunicodechar{≡}{\ensuremath{\mathnormal\equiv}}
\newunicodechar{∀}{\ensuremath{\mathnormal\forall}}
\newunicodechar{∙}{\ensuremath{\mathnormal\bullet}}
\newunicodechar{≣}{\ensuremath{\mathnormal\superequiv}}
\newunicodechar{▿}{\ensuremath{\mathnormal\triangledown}}
\newunicodechar{≤}{\ensuremath{\mathnormal\leq}}
\newunicodechar{↔}{\ensuremath{\mathnormal\leftrightarrow}}
\newunicodechar{⊂}{\ensuremath{\mathnormal\subset}}
\newunicodechar{∘}{\ensuremath{\mathnormal\circ}}
\newunicodechar{∃}{\ensuremath{\mathnormal\exists}}
\newunicodechar{↓}{\ensuremath{\mathnormal\downarrow}}
\newunicodechar{↑}{\ensuremath{\mathnormal\uparrow}}
\newunicodechar{⊔}{\ensuremath{\mathnormal\sqcup}}
\newunicodechar{⊢}{\ensuremath{\mathnormal\vdash}}
\newunicodechar{◇}{\ensuremath{\mathnormal\diamond}}
\newunicodechar{⊙}{\ensuremath{\mathnormal\odot}}
\newunicodechar{⊤}{\ensuremath{\mathnormal\top}}
\newunicodechar{⊥}{\ensuremath{\mathnormal\bot}}
\newunicodechar{∣}{\ensuremath{\mathnormal\mid}}
\newunicodechar{′}{\ensuremath{^\prime}}
\newunicodechar{∅}{\ensuremath{\emptyset}}
\newunicodechar{≺}{\ensuremath{\mathnormal{\prec}}}
\newunicodechar{≼}{\ensuremath{\mathnormal{\preceq}}}
\newunicodechar{∩}{\ensuremath{\mathnormal{\cap}}}
\newunicodechar{⟪}{\ensuremath{\langle\kern-.2em\langle}}
\newunicodechar{⟫}{\ensuremath{\rangle\kern-.2em\rangle}}
\newunicodechar{⊆}{\ensuremath{\mathnormal{\subseteq}}}
\newunicodechar{⊇}{\ensuremath{\mathnormal{\supseteq}}}
\newunicodechar{▻}{\ensuremath{\mathnormal\vartriangleright}}
\newunicodechar{∷}{\ensuremath{::}}
\newunicodechar{►}{\ensuremath{\mathnormal{\blacktriangleright}}}
\newunicodechar{▹}{\ensuremath{\mathnormal{\triangleright}}}
\newunicodechar{□}{\ensuremath{\mathnormal{\square}}}
\newunicodechar{⋯}{\ensuremath{\mathnormal{\ldots}}}
\newunicodechar{▣}{\ensuremath{\mathnormal{\ldots}}}
\newunicodechar{⋮}{\ensuremath{\mathnormal{\quad\quad\vdots}}}
\newunicodechar{∈}{\ensuremath{\mathnormal{\in}}}
\newunicodechar{⊑}{\ensuremath{\mathbin{\sqsubseteq}}}

\newunicodechar{𝓑}{\ensuremath{\mathnormal{\gg\!\!=}}}
\newunicodechar{𝓒}{\ensuremath{\mathcal{C}}}
\newunicodechar{𝓓}{\ensuremath{\mathcal{D}}}
\newunicodechar{𝓡}{\ensuremath{\mathcal{R}}}
\newunicodechar{𝓢}{\ensuremath{\mathcal{S}}}
\newunicodechar{𝓣}{\ensuremath{\mathcal{T}}}
\newunicodechar{𝔃}{\ensuremath{{\color{csy}\mathbf{z}}}}
\newunicodechar{𝓾}{\ensuremath{{\color{csy}\mathbf{u}}}}
\newunicodechar{𝓲}{\ensuremath{{\color{csy}\mathbf{i}}}}
\newunicodechar{𝓴}{\ensuremath{{\color{csy}\mathbf{k}}}}
\newunicodechar{𝓼}{\ensuremath{{\color{csy}\mathbf{s}}}}
\newunicodechar{∶}{$:$}
\newunicodechar{∗}{\ensuremath{\ast}}

\newunicodechar{★}{\ensuremath{\star}}
\newunicodechar{↝}{\ensuremath{\rightsquigarrow}}
\newunicodechar{⇒}{\ensuremath{\Rightarrow}}
\newunicodechar{↦}{\ensuremath{\mapsto}}
\newunicodechar{⊗}{\ensuremath{\otimes}}
\newunicodechar{⊕}{\ensuremath{\oplus}}
\newunicodechar{≃}{\ensuremath{\simeq}}

\newunicodechar{φ}{\ensuremath{\phi}}
\newunicodechar{Φ}{\ensuremath{\Phi}}
\newunicodechar{μ}{\ensuremath{\mu}}
\newunicodechar{σ}{\ensuremath{\sigma}}
\newunicodechar{ξ}{\ensuremath{\xi}}
\newunicodechar{Ξ}{\ensuremath{\Xi}}
\newunicodechar{λ}{\ensuremath{\lambda}}
\newunicodechar{ε}{\ensuremath{\epsilon}}
\newunicodechar{Σ}{\ensuremath{\Sigma}}
\newunicodechar{Δ}{\ensuremath{\Delta}}
\newunicodechar{Π}{\ensuremath{\Pi}}
\newunicodechar{Γ}{\ensuremath{\Gamma}}
\newunicodechar{η}{\ensuremath{\eta}}
\newunicodechar{ζ}{\ensuremath{\zeta}}
\newunicodechar{δ}{\ensuremath{\delta}}
\newunicodechar{α}{\ensuremath{\alpha}}
\newunicodechar{τ}{\ensuremath{\tau}}
\newunicodechar{Λ}{\ensuremath{\Lambda}}
\newunicodechar{ψ}{\ensuremath{\psi}}
\newunicodechar{Ψ}{\ensuremath{\Psi}}
\newunicodechar{π}{\ensuremath{\pi}}
\newunicodechar{θ}{\ensuremath{\theta}}
\newunicodechar{β}{\ensuremath{\beta}}
\newunicodechar{γ}{\ensuremath{\gamma}}
\newunicodechar{ω}{\ensuremath{\omega}}
\newunicodechar{κ}{\ensuremath{\kappa}}
\newunicodechar{ι}{\ensuremath{\iota}}

\newunicodechar{₀}{$_{0}$}
\newunicodechar{₁}{$_{1}$}
\newunicodechar{₂}{$_{2}$}
\newunicodechar{₃}{$_{3}$}
\newunicodechar{₄}{$_{4}$}
\newunicodechar{₅}{$_{5}$}
\newunicodechar{₆}{$_{6}$}
\newunicodechar{₇}{$_{7}$}
\newunicodechar{₈}{$_{8}$}
\newunicodechar{₉}{$_{9}$}
\newunicodechar{¹}{$^{1}$}
\newunicodechar{⁻}{$^{-}$}
\newunicodechar{ᴬ}{$^{A}$}
\newunicodechar{ᴮ}{$^{B}$}
\newunicodechar{ˣ}{$^{×}$}
\newunicodechar{ᶠ}{$^{F}$}
\newunicodechar{ˡ}{$^l$}
\newunicodechar{ʳ}{$^r$}
\newunicodechar{ᴰ}{$^D$}
\newunicodechar{ᵇ}{$^b$}
\newunicodechar{ᵐ}{$_m$}
\newunicodechar{ⁿ}{$_n$}
\newunicodechar{ₙ}{$_n$}
\newunicodechar{ₘ}{$_m$}
\newunicodechar{ᵛ}{$_v$}
\newunicodechar{ᵍ}{$_f$}
\newunicodechar{ᵢ}{$_i$}
\newunicodechar{ₗ}{$_l$}
\newunicodechar{ᵣ}{$_r$}
\newunicodechar{ᶜ}{$^c$}

\newcommand{\kw}[1]{{\color{violet}\textsf{\textbf{#1}}}}
\newcommand{\ks}[1]{{\color{violet}\bm{#1}}}

\newcommand{\KI}[4]{#1 \mid #2 \vdash #3 : #4}
\newcommand{\SC}[2]{#1 \vdash #2}
\newcommand{\TY}[3]{#1 \vdash #2 : #3}

\newcommand{\ZERO}{\mathbb{0}}
\newcommand{\ONE}{\mathbb{1}}

\newcommand{\LET}[4]{\kw{let}\ (#1 : #2) = #3\ \kw{in}\ #4}

\newcommand{\IN}{\kw{in}}
\newcommand{\OUT}{\kw{unin}}
\newcommand{\ARROW}[3]{#1\ {\color{blue} \overset{#2}{\longrightarrow}}\ #3}
\newcommand{\MAP}[2]{\kw{map}\ks{\langle}#2\ks{\rangle}^{#1}}
\newcommand{\FOLD}[2]{\ks{\llparenthesis}\ #2\ \ks{\rrparenthesis}^{#1}}
\newcommand{\FST}{\ks{\pi_{1}}}
\newcommand{\SND}{\ks{\pi_{2}}}
\newcommand{\FORK}[2]{#1\ \ks{\blacktriangle}\ #2}

\newcommand{\INL}{\ks{\iota_1}}
\newcommand{\INR}{\ks{\iota_2}}
\newcommand{\JOIN}[2]{#1\ \ks{\blacktriangledown}\ #2}

\newcommand{\TOP}{\kw{tt}}
\newcommand{\BOT}{\kw{absurd}}

\newcommand{\mcC}{\mathcal{C}}

\newcommand{\D}{\mathcal{D}}
\newcommand{\E}{\mathcal{E}}
\newcommand{\CAT}{\mathbf{CAT}}
\newcommand{\OP}[1]{#1^{\mathsf{op}}}

\newcommand{\cur}[1]{\bm{\uparrow}\!\!(#1)}
\newcommand{\uncur}[1]{\bm{\downarrow}\!\!(#1)}

\newcommand{\hl}[1]{\colorbox{lightgray}{\ensuremath{#1}}}

\newcommand{\SetZ}{\textsc{Set}}
\newcommand{\Set}[1]{\textsc{Set}_{#1}}

\begin{document}

\title{Types and Semantics for Extensible Data Types (Extended Version) 
 }


\author{Cas van der Rest\inst{1}  \and
Casper Bach Poulsen\inst{1}}

\authorrunning{C. van der Rest and C. Bach Poulsen}

\institute{	  
$^{1}$Delft University of Technology, Delft, The Netherlands\\
\email{\{c.r.vanderrest,c.b.poulsen\}@tudelft.nl}}

\maketitle              

\begin{abstract}
  Developing and maintaining software commonly requires (1) adding new data type
  constructors to existing applications, but also (2) adding new functions that
  work on existing data.  Most programming languages have native support for
  defining data types and functions in a way that supports either (1) or (2),
  but not both.  This lack of native support makes it difficult to use and
  extend libraries.  
  A theoretically well-studied solution is to define data types and functions
  using \emph{initial algebra semantics}.  While it is possible to encode this
  solution in existing programming languages, such encodings add syntactic and
  interpretive overhead, and commonly fail to take advantage of the map and fold
  fusion laws of initial algebras which compilers could exploit to generate more
  efficient code.  A solution to these is to provide native support for initial
  algebra semantics.  In this paper, we develop such a solution and present a
  type discipline and core calculus for a language with native support for
  initial algebra
  semantics.  
\end{abstract}

\section{Introduction}
\label{sec:introduction}

A common litmus test for a programming language's capability for modularity is
whether a programmer is able to extend existing data with new ways to construct
it as well as to add new functionality for this data. All in a way that
preserves static type safety; a conundrum which
Wadler~\cite{wadler1998expression} dubbed the \emph{expression problem}. When
working in pure functional programming languages, another modularity question is
how to model side effects modularly using, e.g.,
\emph{monads}~\cite{DBLP:journals/iandc/Moggi91}. Ideally, we would keep the
specific monad used to model the effects of a program abstract and program
against an \emph{interface} of effectful operations instead, defining the syntax
and implementation of such interfaces separately and in a modular fashion.

The traditional approach for tackling these modularity questions in pure
functional programming languages is by embedding the \emph{initial algebra
  semantics}~\cite{goguen1976intial} of inductive data types in the language's
type system. By working with such embeddings in favor of the language's built-in
data types we gain modularity without sacrificing type safety. This approach was
popularized by Swierstra's \emph{Data Types \`{a} la
  Carte}~\cite{DBLP:journals/jfp/Swierstra08} as a solution to the expression
problem, where it was used to derive modular interpreters for a small expression
language. In later work, similar techniques were applied to define the syntax
and implementation of a large class of monads using (algebraic) effects and
handlers based on different flavors of inductively defined \emph{free
  monads}. This was shown to be an effective technique for modularizing both
first order~\cite{DBLP:conf/icfp/KammarLO13} and higher
order~\cite{DBLP:conf/haskell/WuSH14,DBLP:journals/pacmpl/PoulsenR23,DBLP:conf/aplas/BergSPW21}
effectful computations.

The key idea that unifies these techniques is the use of \emph{signature
  functors}, which act as a de facto syntactic representation of an inductive
data type or inductively defined free monad. Effectively, this defines a generic
inductive data type or free monad that takes its constructors as a parameter.
The crucial benefit of this setup is that we can compose data types and effects
by taking the coproduct of signature functors, and we can compose function cases
defined over these signature functors in a similarly modular way.  Inductive
data types and functions in mainstream functional programming languages
generally do not support these kinds of composition.

While embedding signature functors has proven itself as a tremendously useful
technique for enhancing functional languages with a higher degree of type safe
modularity, the approach has some downsides:

\begin{itemize}

\item Encodings of a data type's initial algebra semantics lacks the syntactic
  convenience of native data types, especially when it comes to constructing and
  pattern matching on values. Further overhead is introduced by their limited
  interoperability, which is typically relies on user-defined isomorphisms.
  
\item The connection between initial algebra semantics encodings of data types,
  and the mathematical concepts that motivate them remains implicit. This has
  two drawbacks: (1) the programmer has to write additional code witnessing that
  their definitions possess the required structure (e.g., by defining instances
  of the \texttt{Functor} typeclass), and (2) a compiler cannot leverage the
  properties of this structure, such as by implementing (provably correct)
  optimizations based on the well-known map and fold fusion laws.
  
\end{itemize}

\noindent
In this paper, we explore an alternative perspective by making type-safe
modularity part of the language's design, by including built-in primitives for
the functional programmer's modularity toolkit---e.g., functors, folds,
fixpoints, etc. We believe that this approach has the potential to present the
programmer with more convenient syntax for working with extensible data types
(see, for example, the language design proposed by Van der Rest and Bach
Poulsen~\cite{DBLP:conf/sfp/RestP22}). Furthermore, by supporting type-safe
modularity through dedicated language primitives, we open the door for compilers
to benefit from their properties, for example by applying fusion based
optimizations.

\subsection{Contributions}

The semantics of (nested) algebraic data types has been studied extensively in
the literature (e.g., by Johann et
al.~\cite{DBLP:conf/fossacs/JohannGJ21,DBLP:conf/lics/JohannP19,DBLP:journals/lmcs/JohannG21},
and Abel et
al.~\cite{DBLP:conf/types/AbelM02,DBLP:conf/fossacs/AbelMU03,DBLP:journals/tcs/AbelMU05})
resulting in the development of various calculi with the purpose of studying
different aspects of the semantics of programming with algebraic data types.  In
this paper, we build on these works to develop a core calculus that seeks to
distill the essential language features needed for developing programming
languages with built-in support for type-safe modularity while retaining the
same formal foundations. Although the semantic ideas that we build on to develop
our calculus are generally well-known, their application to improving the design
of functional programming languages has yet to be explored in depth. It is still
future work to leverage the insights gained by developing this calculus in the
design of programming language that provide better ergonomics for working with
extensible data types, but we believe the development of a core calculus
capturing the essentials of programming with extensible data types to be a key
step for achieving this goal. To bridge from the calculus presented in this
paper to a practical language design, features such as \emph{smart
  constructors}, \emph{row types}, and \emph{(functor) subtyping} (as employed,
for example, by Morris and McKinna~\cite{DBLP:journals/pacmpl/MorrisM19} and
Hubers and Morris~\cite{10.1145/3607843}) would be essential. We make the
following technical contributions:

\begin{itemize}

\item We show (in \cref{sec:showcase}) how modular functions over algebraic data
  types in the style of Data Types \`{a} la Carte and modular definitions of
  first-order and higher-order (algebraic) effects and handlers based on
  inductively defined free monads can be captured in the calculus.

\item We present (in \cref{sec:calculus}) a formal definition of the syntax and
  type system.
  
\item
  We give (in \cref{sec:semantics}) a categorical semantics for our calculus.

\item We present (in \cref{sec:operational}) an operational semantics for our
  calculus, and discuss how it relates to the categorical semantics.
  
\end{itemize}

\noindent
\cref{sec:related} discusses related work, and \cref{sec:conclusion} concludes. 
\section{Programming with Extensible Data Types, by Example}
\label{sec:showcase}

The basis of our calculus is the polymorphic $λ$-calculus extended with kinds
and restricted to rank-1 polymorphism, allowing the definition of many familiar
polymorphic functions, such as $(\mathit{id} : ∀α. α ⇒ α) = λ x . x$ or
$(\mathit{const} : ∀α.∀β.α ⇒ β ⇒ α) = λ x . λ y . x$.  Types are closed under
products and coproducts, with the unit type ($\ONE$) and empty type
($\ZERO$) acting as their respective units. Furthermore, we include a type-level
fixpoint ($\mu$), which can be used to encode many well-known algebraic data
types. For example, the familiar type of lists is encoded as
$\mathit{List} \triangleq λα. μ(λX . \ONE + (α × X))$. A key feature of the
calculus is that all higher-order types (i.e., that have one or more type
argument) are, by construction, functorial in all their arguments. While this
imposes some restrictions on the types we can define, it also means that the
programmer gets access to primitive mapping and folding operations that they
would otherwise have to define themselves. For the type $\mathit{List}$, for
example, this means that we get both the usual mapping operation transforming
its elements, as well as an operation corresponding to Haskell's
$\mathit{foldr}$, for free.

Although the mapping and folding primitives for first-order type constructors
(i.e., those taking arguments of kind $★$ and producing a type of kind $★$) are
already enough to solve the expression problem for regular algebraic data types
(\cref{sec:modular-interpreters}) and to encode modular algebraic effects
(\cref{sec:algebraic-effects}), they can readily be generalized to higher-order
type constructors. That is, type constructors that construct higher-order types
from higher-order types. The benefit of this generalization is that our calculus
can also capture the definition of so-called \emph{nested data
  types}~\cite{DBLP:conf/mpc/BirdM98}, which arise as the fixpoint of a
\emph{higher-order functor}. We make essential use of the calculus' higher-order
capabilities in \cref{sec:ho-effects} to define modular handlers for scoped
effects~\cite{DBLP:conf/esop/YangPWBS22} and modular elaborations for
higher-order effects~\cite{DBLP:journals/pacmpl/PoulsenR23}, as in both cases
effect trees that represents monadic programs with higher-order operations is
defined as a nested data type.

\paragraph{Notation.} All code examples in this section directly correspond to
programs in our calculus, but we take some notational liberty to simplify the
exposition. Abstraction and application of type variables is left
implicit. Similarly, we omit first-order universal quantifications. By
convention, we denote type variables bound by type-level $λ$-abstraction using
capital letters (e.g., $X$), and those bound by universal quantification using
Greek letters (e.g., $α$,$β$).

\subsection{Modular Interpreters in the style of Data Types \`{a} la  Carte}
\label{sec:modular-interpreters}

We consider how to define a modular interpreter for a small expression language
of simple arithmetic operations. For starters, we just include literals and
addition. The corresponding BNF equation and signature functor are given below:
\begin{equation*}
  e ::= ℕ \mid e + e 
  \quad\quad\quad\quad\quad
  \mathit{Expr} \triangleq λ X . ℕ + (X × X)
\end{equation*}

\noindent
Now, we can define an $\mathit{eval}$ that maps expressions---given by
the fixpoint of $\mathit{Expr}$---to their result:
\begin{center}
  \vspace{-1.5em}
  \begin{minipage}{0.49\textwidth}
  \begin{align*}
      & \mathit{expr} : ℕ + (ℕ × ℕ) ⇒ ℕ \\
      & \mathit{expr} = \JOIN{(λ x . x)}{(λ x . \FST\ x + \SND\ x)} 
    \end{align*}
  \end{minipage}
  \begin{minipage}{0.49\textwidth}
    \begin{align*} 
      & \mathit{eval} : μ(\mathit{Expr}) ⇒ ℕ \\
      & \mathit{eval} = \FOLD{\mathit{Expr}}{\mathit{expr}}
    \end{align*}
  \end{minipage}
\end{center}

\noindent
Terms typeset in $\kw{purple}$ are built-in operations. $\FST{}$ and $\SND{}$
are the usual projection functions for products, and $\JOIN{-}{-}$ is an
eliminator for coproducts. Following Meijer et
al.~\cite{DBLP:conf/fpca/MeijerFP91}, we write $\FOLD{τ}{\mathit{alg}}$ (i.e.,
``banana brackets'') to denote a fold over the type $μ(τ)$ with an
\emph{algebra} of type $\mathit{alg} : τ\ τ' ⇒ τ'$. The calculus does not
include a general term level fixpoint; the only way to write a function that
recurses on the substructures of a $μ$-type is by using the built-in folding
operation.  While this limits the operations we can define for a given type, it
also ensures that all well-typed terms in the calculus have a well-defined
semantics.

Now, we can extend this expression language with support for a multiplication
operation as follows, where $\mathit{Mul} \triangleq λ X . X × X$:
\begin{center}
  \vspace{-1.5em}
  \begin{minipage}{0.4\textwidth}
    \begin{align*}
      & \mathit{mul} : ℕ × ℕ ⇒ ℕ \\
      & \mathit{mul} = λ x . \FST\ x * \SND\ x 
  \end{align*}
  \end{minipage}
  \begin{minipage}{0.59\textwidth}
    \begin{align*} 
      & \mathit{eval} : μ(\mathit{Expr + Mul}) ⇒ ℕ \\
      & \mathit{eval} = \FOLD{\mathit{Expr + Mul}}{\JOIN{\mathit{expr}}{\mathit{mul}}}
    \end{align*}
  \end{minipage}
\end{center}

\subsection{Modular Algebraic Effects using the Free Monad}
\label{sec:algebraic-effects}

As our second example we consider how to define modular algebraic effects and
handlers~\cite{DBLP:conf/esop/PlotkinP09} in terms of the free monad following Swierstra~\cite{DBLP:journals/jfp/Swierstra08}.  First, we
define the $\mathit{Free}$ type which constructs a free monad for a given
signature functor $f$. We can think of a term with type $\mathit{Free}\ f\ α$ as
a syntactic representation of a monadic program producing a value of type $α$
with $f$ describing the operations which we can use to interact with the monadic
context.
\begin{equation*}
\mathit{Free} : (★ ↝ ★) ↝ ★ ↝ ★ \quad \triangleq \quad λf.λα.μ(λX. α + f X)
\end{equation*}

\noindent
Note that the type $\mathit{Free}$ is actually a functor in both its arguments,
and thus there are two ways to ``map over'' a value of type
$\mathit{Free}\ f\ α$; we can transform the values at the leaves using a
function $α ⇒ β$, or the shape of the nodes using a natural transformation
$∀α . f\ α ⇒ g\ α$. The higher order map can be used, for example, for defining
function that reorders the operations of effect trees with a composite
signature.
\begin{align*} 
  & \mathit{reorder} : \mathit{Free}\ (f + g)\ α ⇒ \mathit{Free}\ (g + f)\ α  \\
  & \mathit{reorder} = \MAP{\mathit{Free}}{\JOIN{\INR}{\INL}}
\end{align*}

\noindent
Here, we use higher order instances at kind $★ ↝ ★$ of the coproduct eliminator
$\JOIN{-}{-}$, the coproduct injection functions $\INL$, $\INR$, and the
functorial map operation $\MAP{-}{-}$.

\emph{Effect handlers} can straightforwardly be implemented as folds over
$\mathit{Free}$. In fact, the behavior of a handler is entirely defined by the
algebra that we use to fold over the effect tree, allowing us write a generic
$\mathit{handle}$ function:
\begin{align*} 
  & \mathit{handle} : (α ⇒ β) ⇒ (f\ (\mathit{Free}\ g\ β) ⇒
    \mathit{Free}\ g\ β) ⇒ \mathit{Free}\ (f + g)\ α ⇒ \mathit{Free}\ g\ β \\
  & \mathit{handle} = λh.λi.\FOLD{α + (f X) + (g X)}{\JOIN{(\IN
    \circ \INL \circ h)}{\JOIN{i}{(\IN \circ \INR)}}}
\end{align*}

\noindent
Here, $\IN{}$ is the constructor of a type-level fixpoint ($μ$).  The fold above
distinguishes three cases: (1) pure values, in which case we return it again
using the function $h$; (2) an operation of the signature $f$ which is handled
using the function $i$; or (3) an operation of the signature $g$ which is
preserved by reconstructing the effect tree and doing nothing.

As an example, we consider how to implement a handler for the $\mathit{Abort}$
effect, which has a single operation indicating abrupt termination of a
computation. We define its signature functor as follows:
\begin{equation*}
  \mathit{Abort} : ★ ↝ ★ \quad \triangleq \quad λX. \ONE
\end{equation*}

\noindent
The definition of $\mathit{Abort}$ ignores its argument, $X$, which is the type
of the continuation. After aborting a computation, there is no continuation,
thus the $\mathit{Abort}$ effect does not need to store one. 
A handler for $\mathit{Abort}$ is then defined like so, invoking the generic
$\mathit{handle}$ function defined above: 
\begin{align*} 
  & \mathit{hAbort} : \mathit{Free}\ (\mathit{Abort} + f)\ α ⇒ \mathit{Free}\ f\ (\mathit{Maybe}\ α)  \\
  & \mathit{hAbort} = \mathit{handle}\ \mathit{Just}\ (λx. \IN\ (\INL\ \mathit{Nothing}))
\end{align*}
  
\subsection{Modular Higher-Order Effects}
\label{sec:ho-effects}

To describe the syntax of computations that interact with their monadic context
through higher-order operations---that is, operations whose arguments can
themselves also be monadic computations---we need to generalize the free monad
as follows. 
\begin{equation*}
  \mathit{Prog} : ((★ ↝ ★) ↝ ★ ↝ ★) ↝ ★ ↝ ★ \quad \triangleq \quad λf.μ(λX.λα.α + (f\ X\ α))
\end{equation*}

\noindent
Note that, unlike the $\mathit{Free}$ type, $\mathit{Prog}$ is defined as the
fixpoint of a higher-order functor. This generalization allows for signature
functors to freely choose the return type of continuations. Following Yang et
al.~\cite{DBLP:conf/esop/YangPWBS22}, we use this additional expressivity to
describe the syntax of higher-order operations by nesting continuations. For
example, the following defines the syntax of an effect for exception catching,
that we can interact with by either throwing an exception, or by declaring an
exception handler that first executes its first argument, and only runs the
second computation if an exception was thrown.
\begin{equation*}
  \mathit{Catch} : (★ ↝ ★) ↝ ★ ↝ ★ \quad \triangleq \quad λX.λα. \ONE + (X (X α) × (X (X
  α)) 
\end{equation*}

\noindent
A value of type $\mathit{Prog}\ \mathit{Catch}\ α$ is then a syntactic
representation of a monadic program that can both throw and catch
exceptions. From this syntactic representation we can proceed in two different
ways. The first option is to replace exception catching with an application of
the $\mathit{hAbort}$ handler, in line with Plotkin and
Pretnar's~\cite{DBLP:conf/esop/PlotkinP09} original strategy for capturing
higher-order operations. In recent work, Bach Poulsen and Van der
Rest~\cite{DBLP:journals/pacmpl/PoulsenR23} demonstrated how such abbreviations
can be made modular and reusable by implementing them as algebras over the
$\mathit{Prog}$ type. Following their approach, we define the following
elaboration of exception catching into a first-order effect tree.
\begin{align*} 
  & \mathit{eCatch} : \mathit{Prog}\ \mathit{Catch}\ α ⇒ \mathit{Free}\
    \mathit{Abort}\ α   \\
  & \mathit{eCatch} = \FOLD{α + \mathit{Catch}\ X\ α}{\JOIN{(\IN \circ
    \INL) \\ &\quad\quad\quad\quad}{\JOIN{(\IN \circ \INR) \\ &\quad\quad\quad\quad}{(λx.\mathit{hAbort}\ (\FST\ x) \gg\!\!= \mathit{maybe}
    \ (\mathit{join}\ (\SND\ x))\ \mathit{id})}}}
\end{align*}

\noindent
Here, the applications of monadic bind ($\gg\!\!=$) and $\mathit{join}$ refer to
the monadic structure of $\mathit{Free}$.  Alternatively, we can define a
handler for exception catching directly by folding over the $\mathit{Prog}$
type, following the \emph{scoped effects} approach by Wu et
al.~\cite{DBLP:conf/haskell/WuSH14}:
\begin{align*} 
  & \mathit{hCatch} : \mathit{Prog}\ (\mathit{Catch} + h)\ α ⇒
    \mathit{Prog}\ h\ (\mathit{Maybe}\ α)   \\
  & \mathit{hCatch} = \FOLD{α +(\mathit{Catch}\ X\ α)+ (h\ X\ α)}{\JOIN{(\IN \circ \INL \circ \mathit{Just}) \\ &\quad\quad\quad\quad}{\JOIN{(λx.\IN\ (\INL\
    \mathit{Nothing}))\\ &\quad\quad\quad\quad}{\JOIN{(λx. \FST\ x \gg\!\!= \mathit{maybe}\ (\SND\ x
    \gg\!\!= \mathit{fwd})\ \mathit{id})) \\ &\quad\quad\quad\quad}{(\IN \circ \INR)}}}}  
\end{align*}

\noindent
Where the function $\mathit{fwd}$ establishes that $\mathit{Maybe}$ commutes
with the $\mathit{Prog}$ type in a suitable way: 
\begin{equation*}
\mathit{fwd} : \mathit{Maybe}\ (\mathit{Prog}\ h\ (\mathit{Maybe}\ α)) ⇒ \mathit{Prog}\ h\ (\mathit{Maybe}\ α)
\end{equation*} 

\noindent
That is, we show that $\mathit{Prog}\ h$ is a \emph{modular carrier} for
$\mathit{Maybe}$~\cite{DBLP:conf/haskell/SchrijversPWJ19}.

As demonstrated, our calculus supports defining higher-order effects and their interpretations.
To conveniently sequence higher-order computations we typically also want to use a monadic bind function, such as $\mathnormal{\gg\!\!=} : \mathit{Prog}\ h\ α \to (α \to \mathit{Prog}\ h\ β) \to Prog\ h\ β$.
While it is possible to define monadic bind for \emph{Free} from \cref{sec:algebraic-effects} in terms of a plain fold, defining the monadic bind for \emph{Prog} generally requires a \emph{generalized fold}~\cite{DBLP:journals/fac/BirdP99,DBLP:conf/esop/YangPWBS22}.
Adding this and other recursion principles~\cite{DBLP:conf/fpca/MeijerFP91} to our calculus is future work.


\begin{figure}[t!]
  \begin{center}
    \begin{tabular}{ r c r c l }
                       &       & $α,β,γ,X,Y$ & $\in $ & $\text{String}$  \\ \\
      \textit{Kind}    $\quad$ & $\ni$ & $k$         & $::=  $ & $★ \mid k ↝ k$     \\ 
      \textit{KindEnv} $\quad$ & $\ni$ & $Δ,Φ$       & $::=  $ & $∅ \mid Δ, α:k$ \\ \\
      \textit{Type}    $\quad$ & $\ni$ & $τ$         & $::=  $ & $α \mid X \mid τ\ τ \mid λX.τ \mid μ(τ) \mid τ ⇒ τ$  \\
                       &       &             & $\mid$ & $\ZERO \mid \ONE \mid τ × τ \mid τ + τ$              \\
      \textit{Scheme}  $\quad$ & $\ni$ & $σ$         & $::=  $ & $∀ α . σ \mid τ$    
    \end{tabular}
  \end{center}
  \caption{Type syntax}
  \label{fig:type-syntax}
\end{figure}

\section{The Calculus}
\label{sec:calculus}

The previous section demonstrated how a language with built-in support for
functors, folds, and fixpoints provides support for defining and working with
state-of-the-art techniques for type safe modular programming. In this section
we present a core calculus for such a language.  The basis of our calculus is
the first-order fragment of System $F^{ω}$---i.e., the polymorphic $λ$-calculus
with kinds, where universal quantification is limited to prenex normal form
\`{a} la Hindley-Milner.  Additionally, the syntax of types, defined in
\cref{fig:type-syntax}, includes primitives for constructing recursive types
($μ(-)$), products ($×$) and coproducts ($+$), as well as a unit type ($\ONE$)
and empty type ($\ZERO$). In the definition of the syntax of types, the use of
$∀$-types is restricted by stratifying the syntax into two layers, types and
type schemes. Consequently, our calculus is, by design, \emph{predicative}:
$∀$-types can quantify over types but not type schemes.

The motivation for this predicative design is that it permits a relatively
straightforward categorical interpretation of $∀$-types in terms of \emph{ends}
(see \cref{sec:universal-quantification}). Whereas the restriction of universal
quantification to prenex normal form is usually imposed to facilitate type
inference, our calculus does not support inference in its current form due to
the structural treatment of data types. In a structural setting, inference
requires the reconstruction of (recursive) data type definitions from values,
which is, in general, not possible.

We remark that the current presentation of the type system is
\emph{declarative}, meaning certain algorithmic aspects crucial for type
checking, such as normalization and equality checking of types, are not covered
in the current exposition. Regarding decidability of the type system: our system
is a subset of System $F_\omega$, whose Church-style formulation is decidable
while its Curry-style formulation is not. As such, we expect our type system to
inherit these properties. Since we are restricting ourselves to a predicative
subset of $F_\omega$, we are optimistic that the Curry-style formulation of our
type system will be decidable too, but verifying this expectation is future
work.

\begin{figure}
  \raggedleft\framebox{$\KI{Δ}{Φ}{τ}{k}$}%
  \begin{mathpar}
    \inferrule[K-Var]
    { k:α \in Δ }
    { \KI{Δ}{Φ}{α}{k} }
    \and
    \inferrule[K-Fvar]
    { Φ(X) ↦ k }
    { \KI{Δ}{Φ}{X}{k} }
    \and
    \inferrule[K-App]
    { \KI{Δ}{Φ}{τ_1}{k_1 ↝ k_2} \\
      \KI{Δ}{Φ}{τ_2}{k_1} }
    {\KI{Δ}{Φ}{τ_1\ τ_2}{k_2}}
    \and
    \inferrule[K-Abs]
    { \KI{Δ}{Φ,(X ↦ k_1)}{τ}{k_2} }
    { \KI{Δ}{Φ}{λ X . τ}{k_1 ↝ k_2} }
    \and
    \inferrule[K-Fix]
    { \KI{Δ}{Φ}{τ}{k ↝ k} }
    { \KI{Δ}{Φ}{μ(τ)}{k} }
    \and
    \inferrule[K-Fun]
    { \KI{Δ}{∅}{τ_1}{★} \\
      \KI{Δ}{Φ}{τ_2}{★} }
    { \KI{Δ}{Φ}{τ_1 ⇒ τ_2}{★} }
    \and
    \inferrule[K-Empty]
    { }
    { \KI{Δ}{Φ}{\ZERO}{k} }
    \and
    \inferrule[K-Unit]
    { }    { \KI{Δ}{Φ}{\ONE}{k} }
    \and
    \inferrule[K-Product]
    { \KI{Δ}{Φ}{τ_1}{k} \\
      \KI{Δ}{Φ}{τ_2}{k} }
    { \KI{Δ}{Φ}{τ_1 × τ_2}{k} }
    \and
    \inferrule[K-Sum]
    { \KI{Δ}{Φ}{τ_1}{k} \\
      \KI{Δ}{Φ}{τ_2}{k} }
    { \KI{Δ}{Φ}{τ_1 + τ_2}{k} }
  \end{mathpar}
  \vspace{.5em}\raggedleft\framebox{$\SC{Δ}{σ}$}%
  \vspace{-2em}
  \begin{mathpar}
    \inferrule[SC-Forall]
    { \SC{Δ,(α ↦ k)}{σ} }
    { \SC{Δ}{∀ α . σ} }
    \and
    \inferrule[SC-Type]
    { \KI{Δ}{∅}{τ}{★} }
    { \SC{Δ}{τ} }
  \end{mathpar}
  \caption{Well-formedness rules for types and type schemes}
  \label{fig:kinding-rules}
\end{figure}

\subsection{Well-Formed Types}

Types are well-formed with respect to a kind $k$, describing the arity of a
type's parameters, if it has any. Well-formedness of types is defined using the
judgment $\KI{Δ}{Φ}{τ}{k}$, stating that the type $τ$ has kind $k$ under
contexts $Δ$ and $Φ$. Similarly, well-formedness of type schemes is defined by
the judgment $\SC{Δ}{σ}$, stating that the type scheme $σ$ is well-formed with
respect to the context $Δ$.

Following Johann et al.~\cite{DBLP:conf/fossacs/JohannGJ21}, well-formedness of
types is defined with respect to two contexts, one containing functorial
variables ($Φ$), and one containing variables with mixed variance
($Δ$). Specifically, the variables in the context $Φ$ are restricted to occur
only in \emph{strictly
  positive}~\cite{DBLP:journals/tcs/AbbottAG05,DBLP:conf/colog/CoquandP88}
positions (i.e., they can never appear to the left of a function arrow), while
the variables in $Δ$ can have mixed variance. This restriction on the occurrence
of the variables in $Φ$ is enforced in the well-formedness rule for function
types, \textsc{K-Fun}, which requires that its domain is typed under an empty
context of functorial variables, preventing the domain type from dereferencing
any functorial variables bound in the surrounding context. While it may seem
overly restrictive to require type expressions to be strictly positive---rather
than merely positive---in $Φ$, this is necessary to ensure that $μ$-types, as
well as its introduction and elimination forms, have a well-defined semantics
(see \cref{sec:recursive-types}). Variables in $Φ$ are bound by type-level
$λ$-abstraction, meaning that any type former with kind $k_1 ↝ k_2$ is
functorial in its argument. In contrast, the variables in $Δ$ are bound by
$∀$-quantification.

Products (×), coproducts (+), units ($\ONE$) and empty types ($\ZERO$) can be
constructed at any kind, reflecting the fact that the corresponding
categorical (co)limits can be lifted from $\SetZ$ to its functor categories by
computing them pointwise. This pointwise lifting of these (co)limits to functor
categories is reflected in the $β$ equalities for these type formers (shown in
\cref{fig:type-equivalence}), which allow an instance at kind $k_1 ↝ k_2$, when
applied with a type argument, to be replaced with an instance at kind $k_2$.

The well-formed judgements for types effectively define a (simply typed) type
level $λ$-calculus with base ``type'' $★$. Consequently, the same type has
multiple equivalent representations in the presence of $β$-redexes, raising the
question of how we should deal with type normalization. The approach we adopt
here is to add a non-syntactic conversion rule to the definition of our type
system that permits any well-formed term to be typed under an equivalent type
scheme. \cref{sec:type-equivalence} discusses type equivalence in more detail.


\subsection{Well-Typed Terms}

\cref{fig:term-syntax} shows the term syntax of our calculus. Along with the
standard syntactic forms of the polymorphic $λ$-calculus we include explicit
type abstraction and application, as well as introduction and elimination forms
for recursive types ($\IN$/$\OUT$), products ($\FST$/$\SND$/$\FORK{-}{-}$),
coproducts ($\INL$/$\INR$/$\JOIN{-}{-}$), and the unit ($\TOP$) and empty
($\BOT$) types. Furthermore, the calculus includes dedicated primitives for
mapping ($\MAP{-}{-}$) and folding ($\FOLD{-}{-}$) over a type.  

\begin{figure}
  \begin{center}
    \begin{tabular}{ r c r c l }
                           &       & $x,y$ & $\in $ & $\text{String}$          \\ \\
      \textit{Env} $\quad$ & $\ni$ & $Γ$   & $::=  $ & $∅ \mid Γ, x : σ$ \\
      \textit{Term}$\quad$ & $\ni$ & $M,N$ & $::=  $ & $x \mid M\ N \mid λx.M \mid \LET{x}{σ}{M}{N}$  \\
                           &       &       & $\mid$ & $Λα.M \mid M\ @τ \mid \IN \mid \OUT \mid \MAP{τ}{M} \mid \FOLD{τ}{M}$ \\
                           &       &       & $\mid$ & $\FST \mid \SND \mid \FORK{M}{N} \mid \INL \mid \INR \mid \JOIN{M}{N} \mid \TOP \mid \BOT$ \\
    \end{tabular}
  \end{center}
  \vspace{1em}
  \begin{center}
    \begin{tabular}{r c l r}
      $\ARROW{τ_1}{★}{τ_2}$         & $\triangleq$ & $τ_1 ⇒ τ_2$ & $\quad
                                                                   \text{(Arrow Types)}$ \\ 
      $\ARROW{τ_1}{(k_1\ ↝\ k_2)}{τ_2}$ & $\triangleq$ & $∀ α .\ \ARROW{τ_1\ α}{k_2}{τ_2\ α}$ & \\
      $\mathit{where}$             &  & $\SC{Δ}{\ARROW{τ_1}{k}{τ_2}} \quad \mathit{if} \quad \KI{Δ}{∅}{τ_1,τ_2}{k}$
    \end{tabular}
  \end{center}
  \caption{Term syntax}
  \label{fig:term-syntax}
\end{figure}

\cref{fig:term-syntax} also includes the definition of \emph{arrow
  types}. In spirit of the syntactic notion of natural transformations used by
Abel et
al.~\cite{DBLP:conf/types/AbelM02,DBLP:conf/fossacs/AbelMU03,DBLP:journals/tcs/AbelMU05}
to study generalized (Mendler) iteration, an arrow type of the form
$\ARROW{τ_1}{k}{τ_2}$ (where $τ_1,τ_2 : k$) defines the type of \emph{morphisms}
between the objects that interpret $τ_1$ and $τ_2$. Arrow types are defined by
induction over $k$, since the precise meaning of morphism for any pair of types
depends on their kind. If $k = ★$, then a morphism between $τ_1$ and $τ_2$ is
simply a function type. However, if $τ_1$ and $τ_2$ have one or more type
argument, they are to be interpreted as objects in a suitable functor category,
meaning that their morphisms are natural transformations. This is reflected in
the definition of arrow types, by unfolding an arrow $\ARROW{τ_1}{k}{τ_2}$ to a
$∀$-type that closes over all type arguments of $τ_1$ and $τ_2$, capturing the
intuition that polymorphic functions cor respond to natural
transformations.\footnote{This intuition is made formal by
  \cref{thm:arrow-currying} in \cref{sec:arrow-semantics}.}  For instance,
we would type the inorder traversal of binary trees as
$\mathit{inorder} : \ARROW{\mathit{Tree}}{★ ↝ ★ }{\mathit{List}}$
($\triangleq ∀α . Tree\ α ⇒ List\ α$), describing a natural transformation
between the $\mathit{Tree}$ and $\mathit{List}$ functors.

\begin{figure}[t!]  
  \vspace{.5em}\raggedleft\framebox{$\TY{Γ}{M}{σ}$}%
  \begin{mathpar}
    \inferrule[T-Var]
    { x : σ \in Γ}
    { \TY{Γ}{x}{σ} }
    \and
    \inferrule[T-App]
    { \TY{Γ}{M}{τ_1 ⇒ τ_2} \\
      \TY{Γ}{N}{τ_1} }
    { \TY{Γ}{M N}{τ_2} }
    \and
    \inferrule[T-Abs]
    { \TY{Γ,(x:τ_1)}{M}{τ_2} }
    { \TY{Γ}{λ x . M}{τ_1 ⇒ τ_2} }
    \and
    \inferrule[T-Let]
    { \TY{Γ}{M}{σ_1} \\ 
      \TY{Γ,x : σ_1}{N}{σ_2} }
    { \TY{Γ}{\LET{x}{σ_1}{M}{N}}{σ_2}}
    \and
    \inferrule[T-TypeAbs]
    { \TY{Γ}{M}{σ} \\
       α \notin \mathsf{freevars}(Γ) }
    { \TY{Γ}{Λα.M}{∀ α . σ}}
    \and
    \inferrule[T-TypeApp]
    { \TY{Γ}{M}{∀ α . σ} }
    { \TY{Γ}{M\ @τ}{σ[τ/α]} }
    \and
    \inferrule[T-In]
    {  }
    { \TY{Γ}{\IN}{\ARROW{τ\ μ(τ)}{k}{μ(τ)}} }
    \and
    \inferrule[T-Out]
    {  }
    { \TY{Γ}{\OUT}{\ARROW{μ(τ)}{k}{τ\ μ(τ)}} }
    \and
    \inferrule[T-Map]
    { \TY{Γ}{M}{\ARROW{τ_1}{k_1}{τ_2}} }
    { \TY{Γ}{\MAP{τ}{M}}{\ARROW{τ\ τ_1}{k_2}{τ\ τ_2}} }
    \and
    \inferrule[T-Fold]
    { \TY{Γ}{M}{\ARROW{τ_1\ τ_2}{k}{τ_2}} }
    { \TY{Γ}{\FOLD{τ_1}{M}}{\ARROW{μ(τ_1)}{k}{τ_2}} }
    \and
    \inferrule[T-Fst]
    { }
    { \TY{Γ}{\FST}{\ARROW{τ_1 × τ_2}{k}{τ_1}} }
    \and
    \inferrule[T-Snd]
    { }
    { \TY{Γ}{\SND}{\ARROW{τ_1 × τ_2}{k}{τ_2}} }
    \and
    \inferrule[T-Fork]
    { \TY{Γ}{M}{\ARROW{τ}{k}{τ_1}} \\
      \TY{Γ}{N}{\ARROW{τ}{k}{τ_2}} }
    { \TY{Γ}{\FORK{M}{N}}{\ARROW{τ}{k}{τ_1 × τ_2}} }
    \and
    \inferrule[T-Inl]
    { }
    { \TY{Γ}{\INL}{\ARROW{τ_1}{k}{τ_1 + τ_2}} }
    \and
    \inferrule[T-Inr]
    { }
    { \TY{Γ}{\INR}{\ARROW{τ_2}{k}{τ_1 + τ_2}} }
    \and
    \inferrule[T-Join]
    { \TY{Γ}{M}{\ARROW{τ_1}{k}{τ}} \\
      \TY{Γ}{M}{\ARROW{τ_2}{k}{τ}} }
    { \TY{Γ}{\JOIN{M}{N}}{\ARROW{τ_1 + τ_2}{k}{τ}} }
    \and
    \inferrule[T-Unit]
    { }
    { \TY{Γ}{\TOP}{\ONE} }
    \and
    \inferrule[T-Empty]
    { }
    { \TY{Γ}{\BOT}{\ARROW{\ZERO}{k}{τ}}}
    \and
    \inferrule[T-Conv]
    { \TY{Γ}{M}{σ_1} \\
      σ_1 \equiv σ_2 }
    { \TY{Γ}{M}{σ_2} }
  \end{mathpar}
  \caption{Well-formed terms}
  \label{fig:typing-rules}
\end{figure}

The typing rules are shown in shown in \cref{fig:typing-rules}.  The rules rely
on arrow types for introduction and elimination forms.  For example, Products
can be constructed at any kind (following rule \textsc{K-Product} in
\cref{fig:kinding-rules}), so the rules for terms that operate on these (i.e.,
\textsc{T-Fst}, \text{T-Snd}, and \textsc{\mbox{T-Fork}}) use arrow types at any
kind $k$. Consequently, arrow types should correspond to morphisms in a suitable
category, such that the semantics of a product type and its
introduction/elimination forms can be expressed as morphisms in this
category.

\subsection{Type Equivalence}
\label{sec:type-equivalence}

\begin{figure}[t]
  \begin{minipage}{0.42\textwidth}  
    \begin{center}
    \begin{tabular}{ r c l }
      $(λX.τ_1)\ τ_2  $ & $\equiv$ & $τ_1[τ_2/X]$ \\
      $(λX.τ\ X)      $ & $\equiv$ & $τ$ \\  
      $(τ_1 × τ_2)\ τ $ & $\equiv$ & $(τ_1\ τ) × (τ_2\ τ)$ \\
      $(τ_1 + τ_2)\ τ $ & $\equiv$ & $(τ_1\ τ) + (τ_2\ τ)$ \\
      $\ONE\ τ       $  & $\equiv$ & $\ONE$ \\
      $\ZERO\ τ      $  & $\equiv$ & $\ZERO$
    \end{tabular}
    \end{center}
  \end{minipage}
  \begin{minipage}{0.57\textwidth}
    \begin{center}
      \begin{tabular} { r c l }
        $T$ & $:=$   & $[] \mid T\ τ \mid τ\ T \mid μ(T) \mid T ⇒ τ \mid τ ⇒ T$ \\
            & $\mid$ & $T × τ \mid τ × T \mid T + τ \mid τ + T$      
      \end{tabular}
      \begin{mathpar}
        \inferrule
        { τ_1 \equiv τ_2 }
        { T[\ τ_1\ ] \equiv T[\ τ_2\ ] }
      \end{mathpar}
    \end{center}
  \end{minipage}
\caption{Equational theory for types}
\label{fig:type-equivalence}
\end{figure}

In the presence of type level $λ$-abstraction and application, the same type can
have multiple representations. For this reason, the type system defined in
\cref{fig:typing-rules} includes a non-syntactic conversion rule that allows a
well-typed term to be re-typed under any equivalent type scheme.  The relevant
equational theory for types is defined in \cref{fig:type-equivalence}, and
includes the customary $β$ and $η$ equivalences for $λ$-terms, as well as $β$
rules for product, sum, unit, and empty types. The equations shown in
\cref{fig:type-equivalence} are motivated by the semantic model we discuss in
\cref{sec:semantics}, in the sense that equivalent types are interpreted to
naturally isomorphic functors. The relation is also reflexive and transitive,
motivated by respectively the identity and composition of natural
isomorphisms. Viewing the equalities in \cref{fig:type-equivalence}
left-to-right provides us with a basis for a normalization strategy for types,
which would be required for implementing the type system.

\section{Categorical Semantics}
\label{sec:semantics}

In this section, we consider how to define a categorical semantics for our
calculus, drawing inspiration from the semantics defined by Johann and
Polonsky~\cite{DBLP:conf/lics/JohannP19} and Johann et
al.~\cite{DBLP:conf/fossacs/JohannGJ21,DBLP:journals/lmcs/JohannG21}. To define
this semantics, we must show that each type in our calculus corresponds to a
functor, and that all such functors have initial algebras. In
\cref{sec:fixpoints} we discuss the requirements for these initial algebras to
exist, and argue informally why they should exist for the functors interpreting
our types. Although Johann and Polonsky~\cite{DBLP:conf/lics/JohannP19} present
a detailed argument for the existence of initial algebras of the functors
underlying nested data types, it is still future work to adapt this argument to
our setting.

The general setup of our semantics is to interpret types of kind $★$ as objects
in $\SetZ$ (the category of sets), higher-order types as functors on $\SetZ$,
and type schemes as objects in $\Set{2}$ (the category of large sets). This size
bump is necessary to model the universal quantification over types in type
schemes. Crucially, $\SetZ$ is a \emph{full subcategory} of $\Set{1}$, as
witnessed by the existence of a fully faithful inclusion functor $I$:
\begin{equation*}
\SetZ \overset{I}{\hookrightarrow} \Set{1}
\end{equation*}
\noindent
Assuming cumulative universes (i.e., the collection of all large sets also
includes all small sets), $I$ is just the identity functor. We remark that both
$\SetZ$ and $\Set{1}$ are \emph{complete} and \emph{cocomplete} and
\emph{cartesian closed}. Importantly, since $I$ is fully faithful, the cartesian
closed structure of $\SetZ$ is reflected in $\Set{1}$ for those objects that lie
in the image of $I$.

The subcategory relation between $\SetZ$ and $\Set{1}$ reflects the syntactic
restriction of types to rank-1 polymorphism: all objects in $\SetZ$ can also be
found in $\Set{1}$, but $\Set{1}$ is sufficiently larger than $\SetZ$ that it
also includes objects modelling quantification over objects in $\SetZ$. This
intuition is embodied by fact that every functor
$F : \OP{\mcC} × \mcC \to \Set{1}$, where $\mcC$ is smaller than $\Set{1}$
(which includes $\SetZ$), has an end in $\Set{1}$. This follows from
completeness of $\Set{1}$~\cite[p.~224,~corollary~2]{maclane2013categories}. We
discuss the use of ends for modelling universal quantification in more detail in
\cref{sec:universal-quantification}.

\subsection{Interpreting Kinds and Kind Environments}
\label{sec:kind-semantics}

We associate with each kind $k$ a category whose objects interpret the types of
that kind. The semantics of kinds is defined by induction over $k$, where we map
the base kind $★$ to $\SetZ$, and kinds of the form $k_1 ↝ k_2$ to
the category of functors between their domain and codomain.\footnote{Here,
  $\CAT$ denotes the (very large) category of large categories. Although $\SetZ$
itself is locally small, its functor categories have a large set of morphisms.}
\begin{center}
\begin{tabular}{ r c l r }
  $⟦ -          ⟧$ & $:$ & $\mathit{Kind} \to \CAT$ & \\ 
  $⟦\ ★\        ⟧$ & $=$ & $\SetZ$  &  \\
  $⟦\ k_1 ↝ k_2\ ⟧$ & $=$ & $[\ ⟦\ k_1\ ⟧\ ,\ ⟦\ k_2\ ⟧\ ]$  &  \\ 
\end{tabular}
\end{center}

\noindent
By interpreting types of kind $k_1 ↝ k_2$ as objects in a functor category, we
formalize the intuition that higher-order types correspond to functors. The
semantics of kind contexts is then defined on a per-entry basis, as a chain of
products of the categories that interpret their elements.
\begin{center}
\begin{tabular}{ r c l r }
  $⟦ -         ⟧$ & $:$ & $\mathit{Context} \to \CAT$ & \\ 
  $⟦\ ∅\       ⟧$ & $=$ & $∙$  &  \\
  $⟦\ Δ,α ↦ k\ ⟧$ & $=$ &$ ⟦\ Δ\ ⟧ × ⟦\ k\ ⟧$ &  \\ 
\end{tabular}
\end{center}

\noindent
Here, $∙$ denotes the \emph{trivial category}, which has a single object,
$\ast$, together with its identity morphism, $id_{\ast}$.  It is worth
mentioning that $∙$ and $-×-$, together with the operation of constructing a
functor category, $[-,-]$, imply that $\CAT$ is a cartesian closed category. We
will use this cartesian closed structure to give a semantics to the fragment of
well-formed types that corresponds to the simply-typed $λ$-calculus.

\subsection{Interpreting Types}
\label{sec:type-semantics}

Since a well-formed type $\KI{Δ}{Φ}{τ}{k}$ is intended to be functorial in all
variables in $Φ$, it is clear that its semantics should be a functor over the
category associated with $Φ$ (i.e., $⟦Φ⟧$). But what about the variables in $Δ$,
which can occur both in covariant and contravariant positions? For example, in
the type of the identity function, $∀α.α ⇒ α$, we cannot interpret the
sub-derivation for $α ⇒ α$ as a functor over the category interpreting its free
variables since there would not be a sensible way to define its action on
morphisms due to the negative occurence of $α$. To account for the mixed
variance of universally quantified type variables, we instead adopt a
\emph{difunctorial semantics}, interpreting types as a functor on the product
category $\OP{⟦ Δ ⟧} × ⟦ Δ ⟧$ (similar representations of type expressions with
mixed variance appear, for example, when considering Mendler-style inductive
types~\cite{DBLP:journals/njc/UustaluV99}, or the object calculus semantics by
Glimming and Ghani~\cite{DBLP:journals/entcs/GlimmingG05}). Well-formed types
(left) and type schemes (right) are interpreted as a functors over their
contexts of the following form:
\begin{mathpar}
  ⟦\ \KI{Δ}{Φ}{τ}{k}\ ⟧ : (\OP{⟦ Δ ⟧} × ⟦ Δ ⟧) × ⟦ Φ ⟧ \to ⟦ k ⟧
  \and
  ⟦\ \SC{Δ}{σ}\ ⟧ : \OP{⟦ Δ ⟧} × ⟦ Δ ⟧ \to \Set{1}
\end{mathpar}

\noindent
Ultimately, the goal of this setup is to interpret $∀$-types as \emph{ends} in
$\Set{1}$, which allows us to formally argue that terms that are well-formed
with an arrow type of the form $\ARROW{τ_1}{k}{τ_2}$ (which unfolds to
$∀\bar{α} . τ_1\ \bar{α} ⇒ τ_2\ \bar{α}$) correspond, in a suitable sense, to
the natural transformations between the functors interpreting $τ_1$ and
$τ_2$. Or, put differently, terms with an arrow type define a morphism between
the interpretation of their domain and codomain. We discuss the semantics of
universal quantification further in \cref{sec:universal-quantification}, and
give a more precise account of the relation between arrow types and natural
transformations in \cref{sec:arrow-semantics}.

\begin{figure}
  \begin{align*}
    ⟦ \KI{Δ}{Φ}{α}{τ} ⟧            & = \textsf{\textbf{lookup}}^Δ_α \circ \pi_2 \circ π_1 \\
    ⟦ \KI{Δ}{Φ}{X}{τ} ⟧            & = \textsf{\textbf{lookup}}^Φ_X \circ π_2 \\ 
    ⟦ \KI{Δ}{Φ}{τ_1\ τ_2}{k_2} ⟧   & = \textsf{eval}\ ∘\ ⟨\ ⟦ \KI{Δ}{Φ}{τ_1}{k_1 ↝ k_2} ⟧ , ⟦ \KI{Δ}{Φ}{τ_2}{k_1} ⟧\ ⟩ \\
    ⟦ \KI{Δ}{Φ}{λX.τ}{k_1 ↝ k_2} ⟧ & = \textsf{curry}({⟦ \KI{Δ}{Φ,X : k_1}{τ}{k_2} ⟧}) \\
    ⟦ \KI{Δ}{Φ}{μ(τ)}{k} ⟧         & = \bm{μ}(⟦ \KI{Δ}{Φ}{τ}{k ↝ k} ⟧) \\
    ⟦ \KI{Δ}{Φ}{τ_1 ⇒ τ_2}{★} ⟧    & = \textsf{\textbf{exp}}(⟦ \KI{Δ}{∅}{τ_1}{★} ⟧ , ⟦ \KI{Δ}{Φ}{τ_2}{★} ⟧ ) \\
    ⟦ \KI{Δ}{Φ}{\ZERO}{★} ⟧        & = \bot \\
    ⟦ \KI{Δ}{Φ}{\ONE}{★}  ⟧        & = \top \\
    ⟦ \KI{Δ}{Φ}{τ_1 × τ_2}{k} ⟧    & = ⟦ \KI{Δ}{Φ}{τ_1}{k} ⟧ × ⟦ \KI{Δ}{Φ}{τ_2}{k} ⟧ \\
    ⟦ \KI{Δ}{Φ}{τ_1 + τ_2}{k} ⟧    & = ⟦ \KI{Δ}{Φ}{τ_1}{k} ⟧ + ⟦ \KI{Δ}{Φ}{τ_2}{k} ⟧ \\
    \\
    ⟦ \SC{Δ}{∀α.σ} ⟧               & = \textsf{\textbf{end}}(\textsf{curry}(⟦
                                     \SC{Δ,α : k}{σ} ⟧ \circ
                                     \textsf{\textbf{sift}}))  \\
    ⟦ \SC{Δ}{τ}    ⟧               & = I \circ ⟦ \KI{Δ}{∅}{τ}{★} ⟧                                 
  \end{align*}
  \caption{Semantics of well-formed types and type schemes}
  \label{fig:type-semantics}
\end{figure}

\cref{fig:type-semantics} defines the semantics of well-formed types and type
schemes. The interpretation of the empty type, unit type, and (co)product types
follow immediately from (co)completeness of $\SetZ$. Since they can be
constructed at any kind, the semantics of (co)product types depends crucially on
the fact that functor categories preserve all (co)limits of their codomain
category, which implies that $⟦ k ⟧$ is (co)complete for any $k$. To interpret
variables, we utilize the cartesian closed structure of $\CAT$ to compute an
appropriate projection based on the position of the variable in the environment.
\begin{center}
\begin{tabular}{ l c l r }
  $\textsf{\textbf{lookup}}^{Δ}_α $ & $:$ & $⟦\ Δ\ ⟧ \to ⟦\ k\ ⟧ $ & \\ 
  $\textsf{\textbf{lookup}}^{Δ,α : k}_{α}$ & $↦$ &  $\pi_2$  &  \\
  $\textsf{\textbf{lookup}}^{Δ,β : k}_{α}$ & $↦$ & $\textsf{\textbf{lookup}}^{Δ}_{α}\ ∘\ \pi_1$  & \quad(where $α \neq β$)\\    
\end{tabular}
\end{center}

\noindent
Similarly, the cartesian closed structure of $\CAT$ also implies the existence
of functors $\textsf{eval} : [\mcC,\D] × \mcC \to \D$ and
$\textsf{curry}(F) : \mcC \to [\D , \E]$, for any $F : \mcC × \D \to \E$, which
immediately provide a semantics for type-level application and abstraction
respectively. The remaining type and type scheme constructors are interpreted
using specifically-defined functors. Although their definitions are typical
examples of how (co)limits are lifted to functor categories by computing them
pointwise, we discuss the definition of these functors separately and in more
detail respectively in \cref{sec:recursive-types} (recursive types),
\cref{sec:function-types} (function types), and
\cref{sec:universal-quantification} ($∀$-types).

\subsubsection{Recursive Types}
\label{sec:recursive-types}

Following the usual categorical interpretation of inductive data
types~\cite{goguen1976intial}, the semantics of recursive types is given by an
\emph{initial algebras}. We summarize the setup here.  An
$F$-\emph{algebra} for an endofunctor $F : \mcC \to \mcC$ is defined as a tuple
$(A, α)$ of an object $A \in C$ (called the \emph{carrier}), and a morphism
$α : FA \to A$. An \emph{algebra homomorphism} between $F$-algebras $(A , α)$
and $(B , β)$ is given by a morphism $f : A \to B$ such that the following
diagram commutes.
\begin{center}
\begin{tikzcd}
FA \arrow[d, "\alpha"'] \arrow[rr, "Ff"] &  & FB \arrow[d, "\beta"] \\
A \arrow[rr, "f"']                       &  & B                    
\end{tikzcd}
\end{center}

\noindent
$F$-algebras and their homomorphisms form a category. If $F$ is an endofunctor,
we denote the initial object of the category of $F$-algebras (which, if it
exists, we refer to as the initial algebra) as $(μF , \textsf{in})$. Initial
algebras give a semantics to inductive data types, with their universal property
providing an induction principle. Given an $F$-algebra $(A , α)$, we denote
unique F-algebra homomorphism that factors through $A$ by
$\textsf{cata}(α) : μF \to A$. Instantiating the diagram above with
$\textsf{cata}(α)$ gives us the familiar universal property of folds,
$\textsf{cata}(α) \circ \textsf{in} = α \circ F(\textsf{cata}(α))$, which
defines their computational behavior.

To interpret recursive types in our calculus, we construct the functor
$\bm{μ}(F)$, which sends objects pointwise to the initial algebras of a functor
$F : \mcC \to [\D , \D]$. For a morphism $f : X \to Y$, the action of
$\bm{μ}(F)$ on $f$ is defined by factoring through the algebra defined by
precomposing the initial algebra of $F(Y)$ with the action of $F$ on $f$, which
defines a natural transformation $F(X) \overset{\cdot}{\to} F(Y)$, at component
$μ(F(Y))$.
\begin{center}
\begin{tabular}{ l c l }
  $\bm{μ}(F)(-)$ & $:$ & $\mcC \to \D$  \\
  $\bm{μ}(F)(x)$ & $↦$ & $μ(F(x))$ \\
  $\bm{μ}(F)(f)$ & $↦$ & $\textsf{cata}(\textsf{in} \circ F(f)_{μ(F(Y))})$ \\
\end{tabular}
\end{center}

\noindent
In general, it is not guaranteed that an initial algebra exists for any
endofunctor $F : \mcC \to \mcC$. Typically, the existence of an initial algebras
is shown by iterating $F$ and showing that it converges, applying the classic
theorem by Ad\'{a}mek~\cite{adamek1974free}. This approach imposes some
additional requirements on the functor $F$ and underlying category $\mcC$, which
we discuss in more detail in \cref{sec:fixpoints}.

\subsubsection{Function Types}
\label{sec:function-types}

The functor $\textsf{\textbf{exp}}(-)$ is defined by mapping onto exponential
objects in $\SetZ$. But we have to take some additional care to ensure that we
can still define its action on morphism, as the polarity of free variables is
reversed in domain of a function type. Indeed, when computed pointwise,
exponential objects give rise to a bifunctor of the form
$\OP{\mcC} \to \mcC \to \mcC$, meaning that functors are not, in general, closed
under exponentiation. To some extent we anticipated this situation already in
the design of our type system by defining the well-formedness rule for function
types such that the context of functorial variables, $Φ$, is discarded in its
domain. Of course, the variables in $Δ$ can occur both in covariant and
contravariant positions, but by adopting a difunctorial semantics we limit
ourselves to a specific class of functors that is closed under
exponentiation. The key observation is that constructing the opposite category
of the product of a category and its opposite is an idempotent (up to
isomorphism) operation. That is, we have the following equivalence of
categories: $\OP{(\OP{C} × \mcC)} ≃ \OP{C} × \mcC$. As a result, a pointwise
mapping of difunctors to exponential objects does give rise to a new difunctor.
We use this fact to our advantage to define the following functor
$\textsf{\textbf{exp}}(F,G)$ for functors $F : \OP{\mcC} × \mcC \to \E$ and
$G : (\OP{C} × \mcC) × \D \to \E$, of which the interpretation of function types
is an instance.
\begin{center}
\begin{tabular}{ l c l }
  $\textsf{\textbf{exp}}(F,G)(-)$          & $:$ & $(\OP{\mcC}×\mcC)×\D \to \E$ \\
  $\textsf{\textbf{exp}}(F,G)((x, y) , z)$ & $↦$ & $G((x,y), z)^{F(y , x)}$ \\
  $\textsf{\textbf{exp}}(F,G)((f, g) , h)$ & $↦$ & $\textsf{curry}(G((f,g),h) \circ \textsf{eval}
                                                   \circ (id_{\textsf{\textbf{exp}}(F,G)((x,y),z)} × F(g,f))$\\ 
\end{tabular}
\end{center}

\noindent
We remark that \textsf{\textbf{exp}(F,G)} does not define an exponential object
in the functor category $[(\OP{\mcC} × \mcC) × \D , \E]$. Fortunately, for
defining the semantics of term level $λ$-abstraction or application it is
sufficient that the action on objects maps to exponentials in $\SetZ$.

\subsubsection{Universal quantification}
\label{sec:universal-quantification}

The semantics of universal quantifications is expressed in terms of ends in the
category $\Set{1}$. If $F : \OP{\mcC} × \mcC \to \D$ is a functor, then an
\emph{end} of $F$ is an object $\int_{x \in \mcC} F(x,x) \in \D$ equipped with a
projection map given by an extranatural transformation
$π_x : \int_{c \in \mcC}F(c,c) \to F(x,x)$. Formally, and end of the $F$ is
defined as the universal wedge of the following diagram:
\begin{equation*} 
F(x,x) \overset{F(id_x,f)}{\longrightarrow} F(x,y) \overset{F(f,id_y)}{\longleftarrow} F(y,y)
\end{equation*} 

\noindent
For all $x,y \in \mcC$ and $f : x \to y$. The universal property of ends then
states that any other wedge $W \in \D$ with maps $i : W \to F(x,x)$ and
$j : W \to F(y,y)$ uniquely factors through $\int_{c \in \mcC} F(c,c)$.
\begin{center}
\begin{tikzcd}
W \arrow[rr, "i"] \arrow[rrd, "j" {yshift=3pt,xshift=-10pt}] \arrow[d, "\textsf{factor}(W)" description, dotted] &  & {F(x,x)} \arrow[rrd, "{F(id_x,f)}"] &  &          \\
{\int_{c \in \mathcal{C}}F(c,c)} \arrow[rru, "\pi_x"' {yshift=-5pt,xshift=-15pt}] \arrow[rr, "\pi_y"']         &  & {F(y,y)} \arrow[rr, "{F(f,id_y)}"'] &  & {F(x,y)}
\end{tikzcd}
\end{center}

\noindent
To model the more general situation where a $∀$-quantified type can contain free
variables that are bound by another quantifier above it in the lexical
hierarchy, we define the semantics of universal quantification in terms of the
\emph{end functor}, $\textsf{\textbf{end}}(-)$, which for a functor
$G : \mcC \to [ \OP{\D} × \D , \E ]$ defines a functor
$\textsf{\textbf{end}}(G) : \mcC \to E$ whose object action is computed
pointwise from ends in $\E$. Its action on morphisms,
$\textsf{\textbf{end}}(f) : \int_{d \in \D}G(X)(d,d) \to \int_{d \in \D}
G(Y)(d,d)$, follows from the universal property stated above.  To
define the action on morphisms, we observe that the object
$\int_{d \in \D} G(X)(d,d)$ is a wedge of the following diagram.
\begin{equation*} 
G(Y)(x,x) \overset{G(Y)(id_x,f)}{\longrightarrow} G(Y)(x,y) \overset{G(Y)(f,id_y)}{\longleftarrow} G(Y)(y,y)
\end{equation*}

\noindent
Where the vertices of the cone are constructed by composing the projection map
with the action of $G$ on $f$, i.e., $G(f)(x,x)\ ∘\ π_x$. By universality, this
wedge uniquely factors through the end $\int_{d \in \D}G(Y)(d,d)$. This
factorization defines the morphism action $\textsf{\textbf{end}(f)}$.
\begin{center}
\begin{tabular}{ l c l}
  $\textsf{\textbf{end}}(G)(-)$ & $:$ & $\mcC \to \E$  \\
  $\textsf{\textbf{end}}(G)(x)$ & $↦$ & $\int_{d \in \D}G(x)(d,d)$ \\
  $\textsf{\textbf{end}}(G)(f)$ & $↦$ & $\textsf{factor}(\int_{d \in \D}G(x)(d,d))$ \\
\end{tabular}
\end{center}

\noindent
An important subtlety here is that $F(X)$ should have an end in $\E$ for every
$X$. In our case, this is a consequence of completeness of
$\Set{1}$.\footnote{See Mac Lane~\cite{maclane2013categories} chapter 9.5
  corollary 2.}  To actually use the functor \textsf{\textbf{end}} to define the
semantics of universal quantifications, we need to precompose the semantics of
its body with the \textsf{\textbf{sift}} functor to separate the quantified
variable from the remainder of the context.
\begin{equation*}
\textsf{\textbf{sift}} : \OP{(⟦ Δ ⟧ × ⟦ k ⟧)} × (⟦ Δ ⟧ × ⟦ k ⟧)) × ⟦ Φ ⟧ \to ((\OP{⟦Δ⟧} × ⟦ Δ ⟧) × ⟦ Φ ⟧) × (\OP{⟦k⟧} × ⟦k⟧)
\end{equation*} 

\noindent
We note that \textsf{\textbf{sift}} defines an isomorphism in $\CAT$. 

\subsection{On the Existence of Initial Algebras}
\label{sec:fixpoints}

In general, it is not the case that any endofunctor has an initial algebra. For
certain classes of endofunctors, it can be shown that an initial algebra exists
by means of Ad\'{a}mek's theorem~\cite{adamek1974free}. Here, we present a
condensed argument for why we expect that functors interpreting well-formed
types of kind $k ↝ k$ (for any $k$) have initial algebras; a more thorough
formal treatment of the construction of initial algebras is a subject of further
study. 

The intuition behind Ad\'{a}mek's construction is that repeated applications of
an endofunctor $F : \mcC \to \mcC$ converge after infinite iterations, reaching
a fixpoint. If $\mcC$ has an initial object and
\emph{$ω$-colimits},\footnote{That is, colimits over diagrams defined as a
  functor on the thin category generated from the poset of natural numbers.}
we can define the initial algebra of $F$ as the $ω$-colimit of the following
chain:
\begin{equation*}
⊥ \overset{!}{\longrightarrow} F⊥ \overset{F!}{\longrightarrow} FF⊥
\overset{FF!}{\longrightarrow} FFF\bot \overset{FFF!}{\longrightarrow} \ldots
\end{equation*} 

\noindent
Where $⊥$ is the initial object in $\mcC$ and $!_X : ⊥ \to X$ the unique map from
$⊥$ to $X$. A crucial stipulation is that $F$ should be \emph{$ω$-cocontinuous},
meaning that it preserves $ω$-colimits.

Thus, for the functors interpreting higher-order types to have an initial
algebra, we must argue that all higher-order types are interpreted to a
$ω$-cocontinuous functor. This prompts a refinement of the semantics for kinds
discussed in \cref{sec:kind-semantics}, where we impose the additional
restriction that the interpretation of a kind of the form $k_1 ↝ k_2$ is a
$ω$-cocontinuous functor from $⟦k_1⟧$ to $⟦k_2⟧$. Subsequently, we must show
that \cref{fig:type-semantics} actually inhabits this refined semantics.

Johann and Polonsky~\cite{DBLP:conf/lics/JohannP19} present an inductive
argument showing the existence of initial algebras for a universe of
higher-kinded data types is similar to our definition of well-formed terms in
\cref{fig:kinding-rules}. While their proof establishes the more general
property of $λ$-cocontinuity (for an arbitrary limit ordinal $λ$) for the
functors interpreting higher-kinded types, we expect that the relevant cases of
their inductive proof---specifically the cases for products, coproducts, type
application, and the $\bm{μ}$ functor---can be adapted to our setting. What
remains is to show that the semantics of type level $λ$-abstraction and function
types is a $ω$-cocontinuous functor. For $λ$-abstraction, we transport along the
currying isomorphism, which should preserve $ω$-cocontinuity. For function
types, we require that the functor $(-)^X : \SetZ \to \SetZ$ is $ω$-cocontinuous
for all $X$, which, as Johann and Polonsky~\cite{DBLP:conf/lics/JohannP19} point
out, is indeed the case. Expanding this proof sketch into a full proof of the
existence of initial algebras is future work.

\subsection{Arrow Types Correspond to Morphisms}
\label{sec:arrow-semantics}

To define the semantics of well-typed terms, it is crucial that we can relate
arrow types---i.e., of the form $\ARROW{τ_1}{k}{τ_2}$---to morphisms in the
category $⟦k⟧$. To make this more precise, consider the typing rule for left
projections. To define its semantics, we would like to use the cartesian
structure of the category $⟦k⟧$, which implies the existence of a
\emph{morphism} $π_1 : ⟦k⟧(x × y , x)$ for $x,y \in ⟦ k ⟧$. However, the rule
\textsc{T-Fst} implies that $\FST$ should be related to an \emph{object} in $\Set{1}$, i.e., $⟦ \ARROW{τ_1 × τ_2}{k}{τ_1} ⟧$. To mediate between
morphisms in $⟦k⟧$ and objects in $\Set{1}$ calls for a suitable currying/uncurrying
isomorphism for arrow types, though we highlight that the required isomorphism
is different from the usual currying isomorphism arising from the existence of
right adjoints for the tensor product in closed monoidal catetegories, in the
sense that $⟦ \ARROW{τ_1}{k}{τ_2} ⟧$ does not define an internal hom for the
objects $⟦ τ_1 ⟧, ⟦ τ_2 ⟧$ but rather inernalizes the morphisms between these
objects in a \emph{different} category.

\begin{theorem}
  \label{thm:arrow-currying}

  Given a kind $k$, morphisms of the category $⟦k⟧$ are internalized as objects
  in $\Set{1}$ through the following bijection between hom-sets:
  \begin{equation}\label{eq:arrow-currying}
  ⟦k⟧(F(\bm{δ}) × ⟦ τ_1 ⟧(\bm{δ}^{\circ}) , ⟦ τ_2 ⟧(\bm{δ})) \quad ≃ \quad \Set{1}(F(\bm{δ}) , ⟦ \ARROW{τ_1}{k}{τ_2} ⟧(\bm{δ}))  
  \end{equation}

  \noindent
  Where $\bm{δ} \in \OP{⟦Δ⟧}×⟦Δ⟧$ and $\bm{δ}^{\circ} \in \OP{(\OP{⟦Δ⟧} × ⟦Δ⟧)}$
  its complement, which is defined by swapping the objects representing
  contravariant respectively covariant occurrences of the variables in $Δ$. Let
  $F : \OP{⟦Δ⟧} × ⟦ Δ ⟧ \to \Set{1}$ be a functor. In a slight abuse of notation, we
  also write $F(\bm{δ})$ for the ``lifting'' of $F$ to an object in the
  (functor) category $⟦k⟧$ that ignores all the additional variables on which
  $⟦τ_1⟧$ and $⟦ τ_2 ⟧$ depend.
  
  \begin{proof}
    We compute the isomorphism as follows, where $k = k_1 ↝ \cdots ↝ k_n ↝ ★$: 
    \begin{equation*}
      \begin{split}
          & ⟦k⟧(F(\bm{δ}) × ⟦ τ_1 ⟧(\bm{δ}^{\circ}) , ⟦ τ_2 ⟧(\bm{δ}) ) \\
        = \quad & \int_{x_1 \in ⟦ k_1 ⟧}\cdots\int_{x_n \in ⟦ k_n ⟧} \Set{1}(F(\bm{δ}) × ⟦ τ_1 ⟧(\bm{δ}^{\circ})(x_1)\cdots(x_n),⟦ τ_2 ⟧(\bm{δ})(x_1)\cdots(x_n))\\
        ≃ \quad  & \int_{x_1 \in ⟦ k_1 ⟧}\cdots\int_{x_n \in ⟦ k_n ⟧} \Set{1}(F(\bm{δ}),⟦ τ_2 ⟧(\bm{δ})(x_1)\cdots(x_n)^{⟦ τ_1 ⟧(\bm{δ}^{\circ})(x_1)\cdots(x_n)}) \\
        ≃ \quad & \Set{1}(F(\bm{δ}),\int_{x_1}\cdots\int_{x_n}⟦ τ_2 ⟧(\bm{δ})(x_1)\cdots(x_n \in ⟦ k_n ⟧)^{⟦ τ_1 ⟧(\bm{δ}^{\circ})(x_1)\cdots(x_n)}) \\
        ≃ \quad & \Set{1}(F(\bm{δ}),⟦ \ARROW{τ_1}{k}{τ_2} ⟧(\bm{δ}))\\   
      \end{split}
    \end{equation*}

    \noindent
    The first step of the derivation rewrites the left-hand side of the
    isomorphism to a sequence of zero or more ends in the category of very large
    sets, allowing us to apply currying for exponentials in $\Set{1}$ in the
    subsequent step. This is justified by cartesian closedness of $\SetZ$, because
    the objects $⟦τ_1⟧(\bm{δ^{\circ}})(x_1)\cdots(x_n)$ and
    $⟦ τ_2 ⟧(\bm{δ})(x_1)\cdots(x_n)$ are included in the image of the fully
    faithful inclusion functor $I$. Next, we use the fact that the
    covariant hom-functor $\Set{1}(x,-)$ is continuous and thus preserves
    ends:\footnote{See Mac Lane~\cite{maclane2013categories}, page 225 Equation
      4.}
    \begin{equation}\label{eq:ends-distr}
      \int_{y \in \mcC} \Set{1}(x , G(y,y)) \quad ≃ \quad \Set{1}(x , \int_{y \in \mcC} G(y,y) ) 
    \end{equation}

    \noindent
    By repeatedly applying the identity above, we can distribute the
    aforementioned sequence of ends over the functor
    $\Set{1}(F(\bm{δ}),-)$. Intuitively, this corresponds to distributing
    universal quantification over logical implication in the scenario that the
    quantified variable does not occur freely in the antecedent, which is
    axiomatized in some flavors of first-order logic, though we apply a much
    more general instance of the same principle here. The final step then
    follows from the standard definition of $η$-equivalence implied by cartesian
    closedness of $\CAT$.
  \end{proof} 
\end{theorem}

\noindent
We write $\cur{-}$/$\uncur{-}$ for the functions that transport along the
\\isomorphism defined in \cref{eq:arrow-currying}. 

\subsection{Interpreting Terms}
\label{sec:term-semantics}

\begin{figure}[t]
  \begin{align*}
    ⟦ \TY{Γ}{x}{σ} ⟧_{\bm{δ}} & = \textbf{\textsf{lookup}}^{Γ}_{x} \\
    ⟦ \TY{Γ}{M\ N}{τ_2} ⟧_{\bm{δ}} & = \textsf{eval} \circ ⟨ ⟦ \TY{Γ}{M}{τ_1 ⇒ τ_2} ⟧_{\bm{δ}} , ⟦ \TY{Γ}{N}{τ_1} ⟧_{\bm{δ}} ⟩  \\
    ⟦ \TY{Γ}{λx.M}{τ_1 ⇒ τ_2} ⟧_{\bm{δ}} & = \textsf{curry}(⟦\TY{Γ,x : τ_1}{M}{τ_2}⟧_{\bm{δ}}) \\
    ⟦ \TY{Γ}{\LET{x}{σ_1}{M}{N}}{σ_2} ⟧_{\bm{δ}} & = \textsf{eval} \circ ⟨ \textsf{curry}(⟦\TY{Γ,x : σ_1}{N}{σ_2}⟧_{\bm{δ}}) , ⟦ \TY{Γ}{N}{σ_1} ⟧_{\bm{δ}} ⟩  \\ 
    ⟦ \TY{Γ}{Λα.M}{∀α.σ} ⟧_{\bm{δ}} & = ⟦ \TY{Γ}{M}{σ} ⟧_{\bm{δ}} \quad \text{(isomorphic per \cref{eq:ends-distr})} \\
    ⟦ \TY{Γ}{M @τ}{σ[τ/α]} ⟧_{\bm{δ}} & = π_{⟦ τ ⟧}\circ ⟦ \TY{Γ}{M}{∀α.σ} ⟧_{\bm{δ}} \\
    ⟦ \TY{Γ}{\IN}{\ARROW{τ\ μ(τ)}{k}{μ(τ)}} ⟧_{\bm{δ}} & = \cur{\textsf{in} \circ \pi_2} \\
    ⟦ \TY{Γ}{\OUT}{\ARROW{μ(τ)}{k}{τ\ μ(τ)}} ⟧_{\bm{δ}} & = \cur{\textsf{unin} \circ \pi_2} \\
    ⟦ \TY{Γ}{\MAP{τ}{M}}{\ARROW{τ\ τ_1}{k_2}{τ\ τ_2}} ⟧_{\bm{δ}} & =
    \ensuremath{\cur{λ(γ,x).⟦τ⟧(\bm{δ})(λy.
    \uncur{⟦\TY{Γ}{M}{\ARROW{τ_1}{k_1}{τ_2}}⟧_{\bm{δ}}}(γ,y))}} \\ 
    ⟦ \TY{Γ}{\FOLD{τ_1}{M}}{\ARROW{μ(τ_1)}{k}{τ_2}} ⟧_{\bm{δ}} & =
    \ensuremath{\cur{λ(γ,x).\textsf{cata}(λy.\uncur{⟦\TY{Γ}{M}{\ARROW{τ_1\
                                                                 τ_2}{k}{τ_2}}⟧_{\bm{δ}}}(γ,y))}} \\ 
    ⟦ \TY{Γ}{\FST}{\ARROW{τ_1 × τ_2}{k}{τ_1}} ⟧_{\bm{δ}} & = \cur{π_1 \circ π_2} \\
    ⟦ \TY{Γ}{\SND}{\ARROW{τ_1 × τ_2}{k}{τ_2}} ⟧_{\bm{δ}} & = \cur{π_2 \circ π_2} \\
    ⟦ \TY{Γ}{\FORK{M}{N}}{\ARROW{τ}{k}{τ_1 × τ_2}} ⟧_{\bm{δ}} & = \cur{\ ⟨\ \uncur{⟦\TY{Γ}{M}{\ARROW{τ}{k}{τ_1}}⟧_{\bm{δ}}}\ ,\ \uncur{⟦\TY{Γ}{N}{\ARROW{τ}{k}{τ_2}}⟧_{\bm{δ}}}\ ⟩\ } \\
    ⟦ \TY{Γ}{\INL}{\ARROW{τ_1}{k}{τ_1 + τ_2}} ⟧_{\bm{δ}} & = \cur{ι_1 \circ \pi_2} \\
    ⟦ \TY{Γ}{\INR}{\ARROW{τ_2}{k}{τ_1 + τ_2}} ⟧_{\bm{δ}} & = \cur{ι_2 \circ \pi_2} \\
    ⟦ \TY{Γ}{\JOIN{M}{N}}{\ARROW{τ_1 + τ_2}{k}{τ}} ⟧_{\bm{δ}} & = \cur{\ [\ \uncur{⟦ \TY{Γ}{M}{\ARROW{τ_1}{k}{τ}} ⟧_{\bm{δ}}}\ ,\ \uncur{⟦ \TY{Γ}{N}{\ARROW{τ_2}{k}{τ}} ⟧_{\bm{δ}}}\ ]\ }\\
    ⟦ \TY{Γ}{\TOP}{\ONE} ⟧_{\bm{δ}} & =\ ! \quad \text{(the unique morphism to the terminal object)} \\
    ⟦ \TY{Γ}{\BOT}{\ZERO ⇒ τ} ⟧_{\bm{δ}} & = \textsf{curry}(h \circ \pi_2) \\
  \end{align*}
  \vspace{-3em}
  \caption{Semantics of Well-Typed Terms.}
  \label{fig:term-semantics}
\end{figure}

Well-typed terms, of the form $\TY{Γ}{M}{σ}$, are interpreted as natural
transformations from the interpretation their context, $⟦Γ⟧$, to the
interpretation of their type, $⟦σ⟧$. At component
$\bm{δ} \in \OP{⟦ Δ ⟧} × ⟦ Δ ⟧$ this transformation is given by a function with
the following type:
\begin{equation*}
  ⟦\ \TY{Γ}{M}{σ}\ ⟧_{\bm{δ}} : ⟦\ Γ\ ⟧(\bm{δ}) \to ⟦\ σ\ ⟧(\bm{δ})  
\end{equation*} 

\noindent
Here, $⟦Γ⟧$ is defined componentwise by mapping contexts to a left-associated
product of its elements, analogous to how we defined the interpretation of kind
contexts in \cref{sec:kind-semantics}. \cref{fig:term-semantics} shows the
interpretation of well-typed terms in its entirety.

The interpretation of $λ$-abstraction and application is defined in terms of the
cartesian closed structure of $\SetZ$, which is preserved by its inclusion in
$\Set{1}$. For a type abstractions of the form $Λα.M$, its semantics follows
from the fact that hom-functors preserves ends (see \cref{eq:ends-distr}), which
implies a bijection between the set of morphisms that interprets the type
abstraction and the set of morphisms into which we interpret its body. We remark
that this only works because $α$ does not occur free in $Γ$, meaning that we
know that $⟦ Γ ⟧$ does not depend on $α$ in
$⟦ \TY{Γ}{M}{σ} ⟧_{\bm{δ},(α , α)} : ⟦ Γ ⟧(\bm{δ},(α , α)) \to ⟦ σ ⟧(\bm{δ},(α ,
α))$, and thus we can view $⟦Γ⟧$ as a constant when applying the
isomorphism. The semantics of a type application $M\ @τ$ is then given by the
projection map at component $⟦τ⟧$ of the end interpreting the type of $M$. For
the introduction and elimination forms of (co)product types, and the unit and
empty type, we define the semantics in terms of the corresponding (co)limits in
$\Set{1}$, applying the currying isomorphism defined in \cref{eq:arrow-currying}
to mediate with arrow types. Similarly, a semantics for the mapping and folding
primitives also follows from the currying isomorphism defined in
\cref{eq:arrow-currying}.

Both the denotation function $⟦-⟧$ as well as the function it computes
are total. Consequently, a well-typed value can be computed from every
well-typed term. In this sense, the categorical model provides us with a sound
computational model of the calculus, which we could implement by writing a
definitional interpreter~\cite{DBLP:journals/lisp/Reynolds98a}. In the next
section, we will discuss how a more traditional small-step operational semantics
can be derived from the same categorical model.

\section{Operational Semantics}
\label{sec:operational}

The previous section gave an overview of a categorical semantics of our
calculus. In this section, we define a small-step operational semantics for our
calculus, and discuss how it relates to the categorical model. 

\subsection{Reduction Rules}

\begin{figure}[ht]
  \begin{center}
    \begin{tabular}{ r c l r }
      $v$ & $:=$ & $λx.M \mid Λα.M \mid \IN\ \hl{\overline{τ}\ v} \mid \OUT\ \hl{\overline{τ}\ v} \mid (\FORK{v_1}{v_2})\ \hl{\overline{τ}\ v} \quad$ & (Values) \\
          & $\mid$ & $ \INL\ \hl{\overline{τ}\ v} \mid \INR\ \hl{\overline{τ}\ v} \mid \MAP{τ′}{v}\ \hl{\overline{τ}} \mid \FOLD{τ′}{v} \hl{\overline{τ}}

                     \mid \FST\ \hl{\overline{τ}} \mid \SND\ \hl{\overline{τ}}$  \\
          & $\mid$ & $(\JOIN{v_1}{v_2})\ \hl{\overline{τ}} \mid \TOP\ \hl{\overline{τ}} \mid \BOT\ \hl{\overline{τ}\ v}$ \\ \\
      $E$ & $:=$   & $[] \mid E\ M \mid v\ E \mid E\ τ \mid \LET{x}{σ}{E}{M} \mid \LET{x}{σ}{v}{E}$ & (Contexts) \\
          & $\mid $& $\MAP{τ}{E} \mid \FOLD{τ}{E} \mid \FORK{E}{M} \mid \FORK{v}{E} \mid \JOIN{E}{M} \mid \JOIN{v}{E}$ 
  \end{tabular}
  \end{center}
  \caption{Values and Evaluation Contexts. \hl{\text{Highlights}} indicate
    optional occurrences of (type) arguments}
  \label{fig:values-contexts}
\end{figure}

We define our operational semantics as a reduction semantics in the style of
Felleisen~and~Hieb~\cite{DBLP:journals/tcs/FelleisenH92}. \cref{fig:values-contexts}
shows the definition of values and evaluation contexts. In our definition of
values, we must account for the fact that language primitives can exist at any
kind. For example, the primitive $\INL$ by itself is a value of type
$\ARROW{τ_1}{k}{τ_1 + τ_2}$. Simultaneously, applying $\INL$ with a value and/or
a sequence of type arguments (the number of which depends on the kind of its
arrow type), also yields a value. In fact, all the \emph{partial applications} of
$\INL$ with only some of its type arguments, or all type arguments but no value
argument, are also values. We use gray highlights to indicate such an optional
application with type and/or value arguments in the definition of values.  

\cref{fig:reduction-rules} defines the reduction rules. We split the rules in
two categories: the first set describes $β$-reduction\footnote{Here, we
  mean ``$β$-reduction'' in the more general sense of simplifying an application
  of an elimination form to an introduction form.} for the various type formers,
while the second set determines how the $\MAP{-}{-}$ primitive
computes. Similar to the definition of values and contexts in
\cref{fig:values-contexts}, we use the notation $\overline{τ}$ to depict a
sequence of zero or more type applications. Unlike for values, these type
arguments are not optional; terms typed by an arrow types must be fully applied
with all their type arguments before they reduce. The notation $N \bullet M$ is
used as a syntactic shorthand for the composition of two arrow types, which is
defined through $η$-expansion of all its type arguments and the term
argument. The reduction rules for the $\MAP{M}{τ}$ primitive are type directed,
in the sense that the selected reduction depends on $τ$. This is necessary,
because in an application of $\MAP{-}{-}$ to a value, there is no way to decide
whether to apply the function or to push the $\MAP{-}{-}$ further inwards by
only looking at the value.

\begin{figure}[ht]
  \begin{center}
  \begin{tabular}{ r c l r }
    ($(λx.M)\ v            $ & $\longrightarrow$ & $M[v/x]$ & (1) \\
    $\LET{x}{σ}{v}{M}     $ & $\longrightarrow$ & $M[v/x]$ & (2) \\
    $(Λα.M)\ τ           $ & $\longrightarrow$ & $M[τ/α]$ & (3) \\
    $\OUT\ \overline{τ}\ (\IN\ \overline{τ}\ v)       $ & $\longrightarrow$ & $v
                                                                              $ & (4) \\
    $\FOLD{τ′}{v_1}\ \overline{τ}\ (\IN\ \overline{τ}\ v_2)$ & $\longrightarrow$ &
                                                                                $v_1\
                                                                               \overline{τ}\
                                                                               (\MAP{τ′}{\FOLD{τ′}{v_1}}\ \overline{τ}\ v_2)$ & (5)  \\
    $\FST\ \overline{τ}\ ((\FORK{v_1}{v_2})\ \overline{τ}\ v)$ &
                                                                   $\longrightarrow$
                                                & $v_1\ \overline{τ}\ v$ & (6) \\
    $\SND\ \overline{τ}\ ((\FORK{v_1}{v_2})\ \overline{τ}\ v)$ &
                                                                   $\longrightarrow$ & $v_2\ \overline{τ}\ v$ & (7)  \\
    $(\JOIN{v_1}{v_2})\ \overline{τ}\ (\INL\ \overline{τ}\ v))$ & $\longrightarrow$ & $v_1\ \overline{τ}\ v$ & (8) \\
    $(\JOIN{v_1}{v_2})\ \overline{τ}\ (\INR\ \overline{τ}\ v))$ & $\longrightarrow$ & $v_2\ \overline{τ}\ v$ & (9) \\
    \\
    $\MAP{(λX.X)}{v_1}\ \overline{τ}\ v_2$ & $\longrightarrow$ & $v_1\ \overline{τ}\ v_2$ & (10) \\
    $\MAP{μ(τ′)}{v_1}\ \overline{τ}\ (\IN\ \overline{τ}\ v_2)$ & $\longrightarrow$ & $\IN\ \overline{τ}\ (\MAP{(τ′\ μ(τ′))}{v_1}\ \overline{τ}\ v_2)$ & (11) \\
    $\MAP{τ_1 × τ_2}{v}\ \overline{τ}\ ((\FORK{v_1}{v_2})\ \overline{τ}\ v_3)$ & $\longrightarrow$ & $(\FORK{(\MAP{τ_1}{v} \bullet v_1)}{(\MAP{τ_2}{v} \bullet v_2)})\ \overline{τ}\ v_3$ & (12) \\
    $\MAP{τ_1 + τ_2}{v_1}\ \overline{τ}\ (\INL\ \overline{τ}\ v_2)$ & $\longrightarrow$ & $\INL\ \overline{τ}\ (\MAP{τ_1}{v_1}\ \overline{τ}\ v_2)$ & (13) \\
    $\MAP{τ_1 + τ_2}{v_1}\ \overline{τ}\ (\INR\ \overline{τ}\ v_2)$ & $\longrightarrow$ & $\INR\ \overline{τ}\ (\MAP{τ_2}{v_1}\ \overline{τ}\ v_2)$ & (14) \\
    $\MAP{\textbf{1}}{v}\ \overline{τ}\ (\TOP\ \overline{τ})$ & $\longrightarrow$ & $\TOP\ \overline{τ}$ & \quad (15) \\ \\
    $N \bullet M$ & $\triangleq$ & $\overline{Λα}.λx.N\ \overline{α}\ (M\ \overline{α}\ x)$ 
  \end{tabular} 
\end{center}
\caption{Reduction rules}
\label{fig:reduction-rules}
\end{figure}

\subsection{Relation to the Denotational Model}

The reduction rules shown in \cref{fig:reduction-rules} define a computational
model for our calculus.  We now discuss how this model arises from the
denotational model discussed in
\cref{sec:semantics}. Informally speaking,
reducing a term should not change its meaning. This intuition is reflected by
the following implication, which states if $M$ reduces $N$, their semantics
should be equal.\footnote{This property implies what Devesas~Campos~and~Levy~\cite{DBLP:conf/fossacs/CamposL18} call \emph{soundness} of the denotational model with respect to the operational model.  Their soundness property is about a big-step relation; ours is small-step.} 

\begin{equation}\label{eq:step-preserves-semantics}
  M \longrightarrow N  \implies ⟦ M ⟧ = ⟦ N ⟧
\end{equation}

\noindent
While we do not give a formal proof of the implication above, by relying on the
categorical model to inform how terms compute we can be reasonably confident
that our semantics does not contain any reductions that violate this
property. That is, all the reductions shown in \cref{fig:reduction-rules} are
supported by an equality of morphisms in the categorical model. 

What does this mean, specifically? The semantics of well-typed terms is given by
a natural transformation, so if $M \longrightarrow N$, $M$ and $N$ should be
interpreted as the same natural transformation. Equivalence of natural
transformations is defined pointwise in terms of the equality relation for
morphisms in the underlying category. In our case, this is the category $\SetZ$,
as terms are interpreted as natural transformations between functors into
$\SetZ$. By studying the properties---expressed as equalities between
morphisms---of the constructions that give a semantics to the different type
formers, and reifying these equalities as syntactic reduction rules, we obtain
an operational model that we conjecture respects the denotational model by
construction.

Let us illustrate this principle with a concrete example. The semantics of a sum
type $τ_1 + τ_2 : k$ is given by a coproduct in the category $⟦ k ⟧$. The
universal property of coproducts tells us that $[ f , g ] \circ ι_1 = f$ and
$[ f , g ] \circ ι_2 = g$, or in other words, constructing and then immediately
deconstructing a coproduct is the same as doing nothing. Rules (8) and (9) in
\cref{fig:reduction-rules} reflect these equations. That is, since the $\INL$,
$\INR$, and $\JOIN{-}{-}$ primitives are interpreted as the injections $ι_1$,
$ι_2$, and unique morphism $[-,-]$ respectively, the universal property of
coproducts tells us that the left-hand side and right-hand side of rule (8) and
(9) in \cref{fig:reduction-rules} are interpreted to equal morphism in the
categorical domain.

The remaining reduction rules are justified by the categorical model in a
similar fashion. More specifically:
\begin{itemize}

  \item Rules (1,2) follow from the $β$-law for exponential objects, which
    states that $\textsf{eval}\ \circ \langle \textsf{curry}(f) , id \rangle = f$.   

  \item Rule (3) holds definitionally, assuming type substitution is
    appropriately defined such that it corresponds to functor application.

  \item Rule (4) follows from Lambek's lemma, which states that the component of
    an initial algebra is always an isomorphism. That is, there exists a
    morphism $\textsf{unin}$ such that $\textsf{unin} \circ \textsf{in} = id$. 

  \item Rule (5) reflects the universal property of folds, i.e.,
    $\textsf{cata}(f) \circ \textsf{in} = f \circ F(\textsf{cata}(f))$. 

  \item Rules (6,7) follow from the universal property of products, which states
    that $\pi_1\ \circ \langle f , g \rangle = f$ and $\pi_2\ \circ \langle f , g \rangle = g$. 

  \item Rule (10) mirrors the identity law for functors, i.e. $F(id) = id$.  

  \item Rule (11) is derived from naturality of the component of the initial
    algebra of higher-order functors, which states that $\bm{μ}(F)(f) \circ
    \textsf{in} = \textsf{in} \circ F(\bm{μ}(F))(f)$. 

  \item Rule (12,13,14,15) are derived from the way (co)-limits are computed
    pointwise in functor categories. For example, the morphism action of the
    product of two functors $F$ and $G$ is defined as
    $(F \times G)(f) = \langle F(f) \circ \pi_1 , G(f) \circ \pi_2 \rangle$, which gives
    rise to rule (12).
  
  \end{itemize}
  

 
\section{Related Work}
\label{sec:related}


%
%

The problem of equipping functional languages with better support for modularity
as been studied extensively in the literature. One of the earlier instances is
the \emph{Algebraic Design Language} (ADL) by Kieburtz and
Lewis~\cite{DBLP:conf/afp/KieburtzL95}, which features language primitives for
specifying computable functions in terms of algebras. ADL overlaps to a large
extent with the first-order fragment of our calculus, but lacks support for
defining nested data types. Zhang et al.~\cite{DBLP:journals/toplas/ZhangSO21}
recently proposed a calculus and language for \emph{compositional programming},
called CP.  Their language design is inspired by \emph{object algebras}, which
in turn is based on the \emph{tagless final}
approach~\cite{DBLP:journals/jfp/CaretteKS09,DBLP:conf/ssgip/Kiselyov10} and
\emph{final algebra semantics}~\cite{DBLP:journals/jcss/Wand79}, which,
according to Wand~\cite[\S{}7]{DBLP:journals/jcss/Wand79}, is an extension of
\emph{initial algebra semantics}.  These lines of work thus provide similar
modularity as initial algebra semantics, but in a way that does not require
\emph{tagged values}.  While the categorical foundations of Zhang et al.'s CP
language seems to be an open question, the language provides flexible support
for modular programming, in part due to its powerful notion of subtyping.  We
are not aware of attempts to model (higher-order) effects and handlers using CP.
In contrast, our calculus is designed to have a clear categorical semantics.
This semantics makes it straightforward to define state of the art type safe
modular (higher-order) effects and handlers. Morris and
McKinna~\cite{DBLP:journals/pacmpl/MorrisM19} define a language that has
built-in support for \emph{row types}, which supports both extensible records
and variants. While their language captures many known flavors of extensibility,
due to parameterizing the type system over a so-called \emph{row theory}
describing how row types behave under composition, rows are restricted
to first order types. Consequently, they cannot describe any modularity that
hinges on the composition of (higher-order) signature functors.

The question of including nested data types in a language's support for
modularity has received some attention as well. For example, Cai et
al.~\cite{DBLP:conf/popl/CaiGO16} develop an extension of $F_ω$ with
equirecursive types tailored to describe patterns from datatype generic
programming. Their calculus is expressive enough to capture the modularity
abstractions discussed in this paper, including those requiring nested data
types, but lacks a denotational model; a correspondence between a subset of
types in their calculus and (traversable) functors is discussed
informally. Similarly, Abel et al.~\cite{DBLP:journals/tcs/AbelMU05} consider an
operational perspective of traversals over nested datatypes by studying several
extensions of $F_ω$ with primitives for \emph{(generalized) Mendler iteration
  and coiteration}. Although these are expressive enough to describe modular
higher-order effects and handlers, their semantic foundation is very different
from the semantics of the primitive fold operation in our calculus. It is future
work to investigate how our calculus can be extended with support for codata. 

A major source of inspiration for the work in this paper are recent works by
Johann and Polonsky~\cite{DBLP:conf/lics/JohannP19}, Johann et
al.~\cite{DBLP:conf/fossacs/JohannGJ21}, and Johann and
Ghiorzi~\cite{DBLP:journals/lmcs/JohannG21}, which respectively study the
semantics and parametricity of nested data types and GADTs. For the latter, the
authors develop a dedicated calculus with a design and semantics that is very
similar to ours. Still, there are some subtle but key differences between the
designs; for example, their calculus does not include general notions of
$∀$-types and function types, but rather integrates these into a single type
representing natural transformations between type constructors. While their
setup does not require the same stratification of the type syntax we adopt here,
it is also slightly less expressive, as the built-in type of transformations is
restricted to closing over 0-arity arguments.

\emph{Data type generic programming} commonly uses a \emph{universe of
  descriptions}~\cite{DBLP:journals/njc/BenkeDJ03}, which is a data type whose
inhabitants correspond to a signature functor. Generic functions are commonly
defined by induction over these descriptions, ranging over a semantic reflection
of the input description in the type system of a dependently-typed host
language~\cite{DBLP:phd/ethos/Dagand13}. In fact, Chapman et
al.~\cite{DBLP:conf/icfp/ChapmanDMM10} considered the integration of
descriptions in a language's design by developing a type theory with native
support for generic programming. We are, however, not aware of any notion of
descriptions that corresponds to our syntax of well-formed types.

\section{Conclusion and Future work}
\label{sec:conclusion}

In this paper, we presented the design and semantics of a calculus with support
for modularity. We demonstrated it can serve as a basis for capturing several
well-known programming patterns for retrofitting type-safe modularity to
functional languages, such as modular interpreters in the style of Data Types
\`{a} la Carte, and modular (higher-order) algebraic effects. The formal
semantics associates these patterns with their motivating concepts, creating the
possibility for a compiler to benefit from their properties such as by
performing fusion-based optimizations.

\subsubsection*{Acknowledgements.}

This research was partially funded by the NWO VENI Composable and
Safe-by-Construction Programming Language Definitions project (VI.Veni.192.259).

\bibliographystyle{splncs04}
\bibliography{references}

\begin{thebibliography}{10}
\providecommand{\url}[1]{\texttt{#1}}
\providecommand{\urlprefix}{URL }
\providecommand{\doi}[1]{https://doi.org/#1}

\bibitem{DBLP:journals/tcs/AbbottAG05}
Abbott, M.G., Altenkirch, T., Ghani, N.: Containers: Constructing strictly
  positive types. Theor. Comput. Sci.  \textbf{342}(1),  3--27 (2005),
  \url{https://doi.org/10.1016/j.tcs.2005.06.002}

\bibitem{DBLP:conf/types/AbelM02}
Abel, A., Matthes, R.: (co-)iteration for higher-order nested datatypes. In:
  Geuvers, H., Wiedijk, F. (eds.) Types for Proofs and Programs, Second
  International Workshop, {TYPES} 2002, Berg en Dal, The Netherlands, April
  24-28, 2002, Selected Papers. Lecture Notes in Computer Science, vol.~2646,
  pp. 1--20. Springer (2002). \doi{10.1007/3-540-39185-1\_1},
  \url{https://doi.org/10.1007/3-540-39185-1\_1}

\bibitem{DBLP:conf/fossacs/AbelMU03}
Abel, A., Matthes, R., Uustalu, T.: Generalized iteration and coiteration for
  higher-order nested datatypes. In: Gordon, A.D. (ed.) Foundations of Software
  Science and Computational Structures, 6th International Conference, {FOSSACS}
  2003 Held as Part of the Joint European Conference on Theory and Practice of
  Software, {ETAPS} 2003, Warsaw, Poland, April 7-11, 2003, Proceedings.
  Lecture Notes in Computer Science, vol.~2620, pp. 54--69. Springer (2003).
  \doi{10.1007/3-540-36576-1\_4},
  \url{https://doi.org/10.1007/3-540-36576-1\_4}

\bibitem{DBLP:journals/tcs/AbelMU05}
Abel, A., Matthes, R., Uustalu, T.: Iteration and coiteration schemes for
  higher-order and nested datatypes. Theor. Comput. Sci.  \textbf{333}(1-2),
  3--66 (2005). \doi{10.1016/j.tcs.2004.10.017},
  \url{https://doi.org/10.1016/j.tcs.2004.10.017}

\bibitem{adamek1974free}
Ad{\'a}mek, J.: Free algebras and automata realizations in the language of
  categories. Commentationes Mathematicae Universitatis Carolinae
  \textbf{15}(4),  589--602 (1974)

\bibitem{DBLP:journals/njc/BenkeDJ03}
Benke, M., Dybjer, P., Jansson, P.: Universes for generic programs and proofs
  in dependent type theory. Nord. J. Comput.  \textbf{10}(4),  265--289 (2003)

\bibitem{DBLP:conf/aplas/BergSPW21}
van~den Berg, B., Schrijvers, T., Poulsen, C.B., Wu, N.: Latent effects for
  reusable language components. In: Oh, H. (ed.) Programming Languages and
  Systems - 19th Asian Symposium, {APLAS} 2021, Chicago, IL, USA, October
  17-18, 2021, Proceedings. Lecture Notes in Computer Science, vol. 13008, pp.
  182--201. Springer (2021),
  \url{https://doi.org/10.1007/978-3-030-89051-3\_11}

\bibitem{DBLP:conf/mpc/BirdM98}
Bird, R.S., Meertens, L.G.L.T.: Nested datatypes. In: Jeuring, J. (ed.)
  Mathematics of Program Construction, MPC'98, Marstrand, Sweden, June 15-17,
  1998, Proceedings. Lecture Notes in Computer Science, vol.~1422, pp. 52--67.
  Springer (1998), \url{https://doi.org/10.1007/BFb0054285}

\bibitem{DBLP:journals/fac/BirdP99}
Bird, R.S., Paterson, R.: Generalised folds for nested datatypes. Formal
  Aspects Comput.  \textbf{11}(2),  200--222 (1999),
  \url{https://doi.org/10.1007/s001650050047}

\bibitem{DBLP:conf/popl/CaiGO16}
Cai, Y., Giarrusso, P.G., Ostermann, K.: System f-omega with equirecursive
  types for datatype-generic programming. In: Bod{\'{\i}}k, R., Majumdar, R.
  (eds.) Proceedings of the 43rd Annual {ACM} {SIGPLAN-SIGACT} Symposium on
  Principles of Programming Languages, {POPL} 2016, St. Petersburg, FL, USA,
  January 20 - 22, 2016. pp. 30--43. {ACM} (2016),
  \url{https://doi.org/10.1145/2837614.2837660}

\bibitem{DBLP:journals/jfp/CaretteKS09}
Carette, J., Kiselyov, O., Shan, C.: Finally tagless, partially evaluated:
  Tagless staged interpreters for simpler typed languages. J. Funct. Program.
  \textbf{19}(5),  509--543 (2009),
  \url{https://doi.org/10.1017/S0956796809007205}

\bibitem{DBLP:conf/icfp/ChapmanDMM10}
Chapman, J., Dagand, P., McBride, C., Morris, P.: The gentle art of levitation.
  In: Hudak, P., Weirich, S. (eds.) Proceeding of the 15th {ACM} {SIGPLAN}
  international conference on Functional programming, {ICFP} 2010, Baltimore,
  Maryland, USA, September 27-29, 2010. pp. 3--14. {ACM} (2010),
  \url{https://doi.org/10.1145/1863543.1863547}

\bibitem{DBLP:conf/colog/CoquandP88}
Coquand, T., Paulin, C.: Inductively defined types. In: Martin{-}L{\"{o}}f, P.,
  Mints, G. (eds.) COLOG-88, International Conference on Computer Logic,
  Tallinn, USSR, December 1988, Proceedings. Lecture Notes in Computer Science,
  vol.~417, pp. 50--66. Springer (1988),
  \url{https://doi.org/10.1007/3-540-52335-9\_47}

\bibitem{DBLP:phd/ethos/Dagand13}
Dagand, P.: A cosmology of datatypes : reusability and dependent types. Ph.D.
  thesis, University of Strathclyde, Glasgow, {UK} (2013),
  \url{http://oleg.lib.strath.ac.uk/R/?func=dbin-jump-full\&object\_id=22713}

\bibitem{DBLP:conf/fossacs/CamposL18}
Devesas~Campos, M., Levy, P.B.: A syntactic view of computational adequacy. In:
  Baier, C., Lago, U.D. (eds.) Foundations of Software Science and Computation
  Structures - 21st International Conference, {FOSSACS} 2018, Held as Part of
  the European Joint Conferences on Theory and Practice of Software, {ETAPS}
  2018, Thessaloniki, Greece, April 14-20, 2018, Proceedings. Lecture Notes in
  Computer Science, vol. 10803, pp. 71--87. Springer (2018),
  \url{https://doi.org/10.1007/978-3-319-89366-2\_4}

\bibitem{DBLP:journals/tcs/FelleisenH92}
Felleisen, M., Hieb, R.: The revised report on the syntactic theories of
  sequential control and state. Theor. Comput. Sci.  \textbf{103}(2),  235--271
  (1992), \url{https://doi.org/10.1016/0304-3975(92)90014-7}

\bibitem{DBLP:journals/entcs/GlimmingG05}
Glimming, J., Ghani, N.: Difunctorial semantics of object calculus. In: Bono,
  V., Bugliesi, M., Drossopoulou, S. (eds.) Proceedings of the Second Workshop
  on Object Oriented Developments, {WOOD} 2004, London, UK, August 30, 2004.
  Electronic Notes in Theoretical Computer Science, vol.~138, pp. 79--94.
  Elsevier (2004), \url{https://doi.org/10.1016/j.entcs.2005.09.012}

\bibitem{goguen1976intial}
Goguen, J.A.: An intial algebra approach to the specification, correctness and
  implementation of abstract data types. IBM Research Report  \textbf{6487}
  (1976)

\bibitem{10.1145/3607843}
Hubers, A., Morris, J.G.: Generic programming with extensible data types: Or,
  making ad hoc extensible data types less ad hoc. Proc. ACM Program. Lang.
  \textbf{7}(ICFP) (aug 2023), \url{https://doi.org/10.1145/3607843}

\bibitem{DBLP:journals/lmcs/JohannG21}
Johann, P., Ghiorzi, E.: Parametricity for nested types and gadts. Log. Methods
  Comput. Sci.  \textbf{17}(4) (2021),
  \url{https://doi.org/10.46298/lmcs-17(4:23)2021}

\bibitem{DBLP:conf/fossacs/JohannGJ21}
Johann, P., Ghiorzi, E., Jeffries, D.: Parametricity for primitive nested
  types. In: Kiefer, S., Tasson, C. (eds.) Foundations of Software Science and
  Computation Structures - 24th International Conference, {FOSSACS} 2021, Held
  as Part of the European Joint Conferences on Theory and Practice of Software,
  {ETAPS} 2021, Luxembourg City, Luxembourg, March 27 - April 1, 2021,
  Proceedings. Lecture Notes in Computer Science, vol. 12650, pp. 324--343.
  Springer (2021), \url{https://doi.org/10.1007/978-3-030-71995-1\_17}

\bibitem{DBLP:conf/lics/JohannP19}
Johann, P., Polonsky, A.: Higher-kinded data types: Syntax and semantics. In:
  34th Annual {ACM/IEEE} Symposium on Logic in Computer Science, {LICS} 2019,
  Vancouver, BC, Canada, June 24-27, 2019. pp. 1--13. {IEEE} (2019),
  \url{https://doi.org/10.1109/LICS.2019.8785657}

\bibitem{DBLP:conf/icfp/KammarLO13}
Kammar, O., Lindley, S., Oury, N.: Handlers in action. In: Morrisett, G.,
  Uustalu, T. (eds.) {ACM} {SIGPLAN} International Conference on Functional
  Programming, ICFP'13, Boston, MA, {USA} - September 25 - 27, 2013. pp.
  145--158. {ACM} (2013), \url{https://doi.org/10.1145/2500365.2500590}

\bibitem{DBLP:conf/afp/KieburtzL95}
Kieburtz, R.B., Lewis, J.: Programming with algebras. In: Jeuring, J., Meijer,
  E. (eds.) Advanced Functional Programming, First International Spring School
  on Advanced Functional Programming Techniques, B{\aa}stad, Sweden, May 24-30,
  1995, Tutorial Text. Lecture Notes in Computer Science, vol.~925, pp.
  267--307. Springer (1995), \url{https://doi.org/10.1007/3-540-59451-5\_8}

\bibitem{DBLP:conf/ssgip/Kiselyov10}
Kiselyov, O.: Typed tagless final interpreters. In: Gibbons, J. (ed.) Generic
  and Indexed Programming - International Spring School, {SSGIP} 2010, Oxford,
  UK, March 22-26, 2010, Revised Lectures. Lecture Notes in Computer Science,
  vol.~7470, pp. 130--174. Springer (2010),
  \url{https://doi.org/10.1007/978-3-642-32202-0\_3}

\bibitem{maclane2013categories}
MacLane, S.: Categories for the Working Mathematician. Springer-Verlag, New
  York (1971), graduate Texts in Mathematics, Vol. 5

\bibitem{DBLP:conf/fpca/MeijerFP91}
Meijer, E., Fokkinga, M.M., Paterson, R.: Functional programming with bananas,
  lenses, envelopes and barbed wire. In: Hughes, J. (ed.) Functional
  Programming Languages and Computer Architecture, 5th {ACM} Conference,
  Cambridge, MA, USA, August 26-30, 1991, Proceedings. Lecture Notes in
  Computer Science, vol.~523, pp. 124--144. Springer (1991),
  \url{https://doi.org/10.1007/3540543961\_7}

\bibitem{DBLP:journals/iandc/Moggi91}
Moggi, E.: Notions of computation and monads. Inf. Comput.  \textbf{93}(1),
  55--92 (1991), \url{https://doi.org/10.1016/0890-5401(91)90052-4}

\bibitem{DBLP:journals/pacmpl/MorrisM19}
Morris, J.G., McKinna, J.: Abstracting extensible data types: or, rows by any
  other name. Proc. {ACM} Program. Lang.  \textbf{3}({POPL}),  12:1--12:28
  (2019), \url{https://doi.org/10.1145/3290325}

\bibitem{DBLP:conf/esop/PlotkinP09}
Plotkin, G.D., Pretnar, M.: Handlers of algebraic effects. In: Castagna, G.
  (ed.) Programming Languages and Systems, 18th European Symposium on
  Programming, {ESOP} 2009, Held as Part of the Joint European Conferences on
  Theory and Practice of Software, {ETAPS} 2009, York, UK, March 22-29, 2009.
  Proceedings. Lecture Notes in Computer Science, vol.~5502, pp. 80--94.
  Springer (2009), \url{https://doi.org/10.1007/978-3-642-00590-9\_7}

\bibitem{DBLP:journals/pacmpl/PoulsenR23}
Poulsen, C.B., van~der Rest, C.: Hefty algebras: Modular elaboration of
  higher-order algebraic effects. Proc. {ACM} Program. Lang.
  \textbf{7}({POPL}),  1801--1831 (2023), \url{https://doi.org/10.1145/3571255}

\bibitem{DBLP:conf/sfp/RestP22}
van~der Rest, C., Poulsen, C.B.: Towards a language for defining reusable
  programming language components - (project paper). In: Swierstra, W., Wu, N.
  (eds.) Trends in Functional Programming - 23rd International Symposium, {TFP}
  2022, Virtual Event, March 17-18, 2022, Revised Selected Papers. Lecture
  Notes in Computer Science, vol. 13401, pp. 18--38. Springer (2022),
  \url{https://doi.org/10.1007/978-3-031-21314-4\_2}

\bibitem{DBLP:journals/lisp/Reynolds98a}
Reynolds, J.C.: Definitional interpreters for higher-order programming
  languages. High. Order Symb. Comput.  \textbf{11}(4),  363--397 (1998),
  \url{https://doi.org/10.1023/A:1010027404223}

\bibitem{DBLP:conf/haskell/SchrijversPWJ19}
Schrijvers, T., Pir{\'{o}}g, M., Wu, N., Jaskelioff, M.: Monad transformers and
  modular algebraic effects: what binds them together. In: Eisenberg, R.A.
  (ed.) Proceedings of the 12th {ACM} {SIGPLAN} International Symposium on
  Haskell, Haskell@ICFP 2019, Berlin, Germany, August 18-23, 2019. pp. 98--113.
  {ACM} (2019), \url{https://doi.org/10.1145/3331545.3342595}

\bibitem{DBLP:journals/jfp/Swierstra08}
Swierstra, W.: Data types {\`{a}} la carte. J. Funct. Program.  \textbf{18}(4),
   423--436 (2008), \url{https://doi.org/10.1017/S0956796808006758}

\bibitem{DBLP:journals/njc/UustaluV99}
Uustalu, T., Vene, V.: Mendler-style inductive types, categorically. Nord. J.
  Comput.  \textbf{6}(3), ~343 (1999)

\bibitem{wadler1998expression}
Wadler, P.: The expression problem. \\
  \url{http://homepages.inf.ed.ac.uk/wadler/papers/expression/expression.txt}
  (1998), accessed: 2020-07-01

\bibitem{DBLP:journals/jcss/Wand79}
Wand, M.: Final algebra semantics and data type extensions. J. Comput. Syst.
  Sci.  \textbf{19}(1),  27--44 (1979),
  \url{https://doi.org/10.1016/0022-0000(79)90011-4}

\bibitem{DBLP:conf/haskell/WuSH14}
Wu, N., Schrijvers, T., Hinze, R.: Effect handlers in scope. In: Swierstra, W.
  (ed.) Proceedings of the 2014 {ACM} {SIGPLAN} symposium on Haskell,
  Gothenburg, Sweden, September 4-5, 2014. pp. 1--12. {ACM} (2014),
  \url{https://doi.org/10.1145/2633357.2633358}

\bibitem{DBLP:conf/esop/YangPWBS22}
Yang, Z., Paviotti, M., Wu, N., van~den Berg, B., Schrijvers, T.: Structured
  handling of scoped effects. In: Sergey, I. (ed.) Programming Languages and
  Systems - 31st European Symposium on Programming, {ESOP} 2022, Held as Part
  of the European Joint Conferences on Theory and Practice of Software, {ETAPS}
  2022, Munich, Germany, April 2-7, 2022, Proceedings. Lecture Notes in
  Computer Science, vol. 13240, pp. 462--491. Springer (2022),
  \url{https://doi.org/10.1007/978-3-030-99336-8\_17}

\bibitem{DBLP:journals/toplas/ZhangSO21}
Zhang, W., Sun, Y., d.~S.~Oliveira, B.C.: Compositional programming. {ACM}
  Trans. Program. Lang. Syst.  \textbf{43}(3),  9:1--9:61 (2021),
  \url{https://doi.org/10.1145/3460228}

\end{thebibliography}

\appendix

\end{document}